\documentclass[a4paper,11pt]{article}
\usepackage{jcappub}

\usepackage{blindtext}
\usepackage{mathtools}
\usepackage{xcolor}
\usepackage{textcomp,gensymb}
\usepackage{booktabs}
\usepackage{mathrsfs}
\usepackage{braket}

\usepackage[utf8]{inputenc}
\usepackage[english]{babel}
\usepackage{hyperref}
\usepackage{multirow}
\usepackage{booktabs}
\definecolor{mypink1}{rgb}{0.858, 0.188, 0.478}
\definecolor{mypink2}{RGB}{219, 48, 122}
\hypersetup{
    colorlinks=true,
    linkcolor=blue,
    filecolor=blue,      
    urlcolor=blue,
    citecolor=blue,
}
\usepackage{color,colortbl}
\definecolor{LightCyan}{rgb}{0.88,1,1}
\usepackage{tensind}
\tensordelimiter{?}

\usepackage{graphicx}
\usepackage{hyperref}
\DeclareGraphicsExtensions{.bmp,.png,.jpg,.pdf}
\usepackage{amsmath}
\usepackage{courier}
\usepackage{verbatim}
\usepackage{amssymb}
\usepackage{amsfonts}
\usepackage{natbib}
\usepackage{soul}

\usepackage{cleveref}

\newcommand{\field}{\boldsymbol{\Phi}}
\newcommand{\vx}{\boldsymbol{x}}
\newcommand{\vy}{\boldsymbol{y}}

\newcommand{\vk}{\boldsymbol{k}}
\newcommand{\vep}{\boldsymbol{\varepsilon}}
\newcommand{\rhoDM}{\rho_\mathrm{DM}}
\newcommand{\de}{\mathrm{d}}
\newcommand{\cY}{\mathcal{Y}}

\newcommand{\GeV}{\mathrm{GeV}}
\newcommand{\cm}{\mathrm{cm}}
\def\bsdexcess{{\footnotesize\textsc{BSD--excess-power}}\xspace}
\def\lpsd{{\footnotesize\textsc{LPSD}}\xspace}
\def\bsdxcorr{{\footnotesize\textsc{BSD--xcorr}}\xspace}
\def\sftxcorr{{\footnotesize\textsc{SFT--xcorr}}\xspace}

\title{Optimising ultra-light dark matter searches with ground-based interferometers}

\author[1]{Paola C.~M.~Delgado,}
\author[2]{Ornella J.~Piccinni,}
\author[1]{and Federico R.~Urban}

\affiliation[1]{CEICO, FZU – Institute of Physics of the Czech Academy of Sciences, Na Slovance 1999/2, 182 00 Prague, Czech Republic}
\affiliation[2]{Universitat de les Illes Balears, IAC3 - IEEC, Cra.\ de Valldemossa km~7.5. 07122 Palma, Spain}

\emailAdd{delgado@fzu.cz}
\emailAdd{ornella.piccinni@uib.es}
\emailAdd{federico.urban@fzu.cz}

\abstract{Ultra-light dark matter fields induce nearly monochromatic signals in gravitational-wave detectors through their coupling to the Standard Model. Their spectral morphology exhibits features caused by sidereal modulation that, for frequencies below $\sim 30~$Hz, enable discrimination between spin-1 and spin-2 ultra-light dark matter signals, provided sufficient signal-to-noise ratio. In the context of LIGO--Virgo--KAGRA search techniques, we show that incorporating these spectral features can improve current excess-power constraints at low frequencies by up to $\sim36\%$. Additionally, we propose an optimised implementation of the cross-correlation statistics within the Band-Sampled-Data framework, enhancing the sensitivity of cross-correlation searches across nearly the entire frequency range, reaching up to $\sim42\%$ at low frequencies and $\sim35\%$ at high frequencies.}

\begin{document}

\maketitle

\tableofcontents

%%%%%%%%%%%%%%%%%%%%%%%%%%%%%%%%%%%%%%%%%%%%
\section{Introduction}
\label{sec:introduction}
%%%%%%%%%%%%%%%%%%%%%%%%%%%%%%%%%%%%%%%%%%%%

Earth-based laser interferometers, LIGO~\cite{LIGOScientific:2014pky}, Virgo~\cite{VIRGO:2014yos} and KAGRA~\cite{KAGRA:2020tym} (LVK) foremost among them, have established gravitational radiation as a new, independent observational channel that gives direct access to the dynamics of compact objects and the strong-field regime of gravity~\cite{LIGOScientific:2016aoc,LIGOScientific:2017vwq,KAGRA:2021vkt}. At the same time, laser interferometers have been proven to be powerful instruments for the direct detection of bosonic ultra-light dark matter~\cite{Guo:2019ker,Grote:2019uvn,Armaleo:2020efr,Miller:2025yyx}, a form of dark matter that, owing to its ultra-light mass and very large occupation number, behaves as a persistent, coherent superposition of nearly monochromatic waves~\cite{Ferreira:2020fam}. Searches for ultra-light dark matter signals with LVK data from the first part of the fourth observing run (O4a)~\cite{PhysRevD.111.062002,2025arXiv250818079T,Soni_2025} have delivered the most stringent constraints on the coupling constants of vector (spin-1) and tensor (spin-2) bosons in most of the frequency range \(10\,\mathrm{Hz} \leq f \leq 2\,\mathrm{kHz}\), corresponding to an ultra-light dark matter mass in the range \(4.13\times 10^{-14}\,\mathrm{eV} \lesssim m \lesssim 8.27\times 10^{-12}\,\mathrm{eV}\)~\cite{LIGOScientific:2025ttj}. The same principle applies to other laser interferometers such as GEO600~\cite{Vermeulen:2021epa} and, at high frequency, the Fermilab Holometer~\cite{Aiello:2021wlp}, both of which place competitive limits on scalar (spin-0) ultra-light dark matter couplings. In the future, the space-based LISA~\cite{Yu:2023iog} (see also~\cite{Miller:2023kkd}) and QUEST~\cite{Patra:2024eke} promise to do the same in the mHz and MHz bands, respectively.

The ultra-light dark matter signal in the LVK detectors is: (a) narrowband, with a dispersion of order \(v_0^2\sim10^{-6}\), where \(v_0=7.67\times 10^{-4} c\) is the virial velocity of the local dark matter halo in natural units; (b) correlated, because the de Broglie wavelength \(\lambda_\mathrm{dB} \doteq 2\pi / (m v_0)\) of the dark matter is much larger than the separation between LVK detectors; (c) stochastic for an integration time longer than the ultra-light dark matter coherence time \(t_\mathrm{coh} \doteq 4\pi / (m v_0^2)\), which itself is typically shorter than standard observing runs~\cite{Centers:2019dyn, Nakatsuka:2022gaf}. Searches in the LVK band have exploited these signal properties using cross-correlation and excess-power methods. In particular, in the latest O4a direct dark matter search, three different methods have been applied~\cite{PhysRevD.110.042001}: the Short Fourier Transform--cross-correlation (\sftxcorr) method~\cite{Pierce:2018xmy,LIGOScientific:2021ffg}, and two excess-power approaches, namely the \bsdexcess~\cite{Miller:2020vsl}, built on top of the Band-Sampled-Data (BSD) libraries~\cite{Piccinni:2018akm}, and the logarithmic power spectral density (\lpsd) method~\cite{TROBS2006120,gottel2025fastprecisespectralanalysis,Gottel_2024}. The \sftxcorr method is designed to capture the correlation between different detectors, thereby enhancing the signal-to-noise ratio (SNR) by suppressing uncorrelated noise. The \bsdexcess
 and \lpsd methods instead tune the search integration time to the expected signal coherence time in each frequency band, in order to contain the full width of the signal for each time segment.

In this work, we propose a new approach that combines the strengths of the cross-correlation statistics and the flexibility of the excess-power methods in order to optimise the sensitivity to the ultra-light dark matter signal. In practice, we develop an improved implementation of the cross-correlation statistics within the BSD framework, allowing for the search integration time to vary with frequency, in such a way that it matches the optimal integration time; we call this new implementation ``BSD--cross-correlation'' (\bsdxcorr). In order to determine the optimal search integration time, we study the ultra-light dark matter signal morphology in the frequency domain, focussing on the spin-1 and spin-2 cases.\footnote{Spin-0 ultra-light dark matter can also be searched for with LVK data; however, existing limits from fifth-force experiments outperform the LVK sensitivity by several orders of magnitude, so we do not pursue this case further~\cite{LIGOScientific:2025ttj}.} We take into account the frequency-dependent broadening of the dark matter spectrum caused by the dark matter velocity distribution and, for the first time, the presence of multiple peaks in the spectrum caused by the sidereal rotation of the Earth, in close analogy with the spectral structure of continuous gravitational waves~\cite{Jaranowski:1998qm}. We show that the structure of the peaks depend on the dark matter spin via the detector response function, thereby offering a new possible way, beyond the overlap reduction functions (ORFs)~\cite{Manita:2023mnc} and the Wiener filter~\cite{Miller:2022wxu}, to distinguish between such signals. The separation of peaks is independent of frequency and it is only clearly visible below \(\sim 30\,\mathrm{Hz}\); nonetheless, it contributes to an additional signal broadening that needs to be included at all frequencies in order to correctly estimate the optimal search integration time. 
Using the new \bsdxcorr approach proposed in this work, we estimate a frequency-dependent sensitivity improvement of up to $42\%$ ($35\%$) at low (high) frequencies relative to the \sftxcorr method. Furthermore, accounting for the additional spectral structure induced by sidereal modulation in excess-power methods yields estimated low-frequency improvements of $36\%$ and $29\%$ for the \bsdexcess and \lpsd methods, respectively.

This paper is organised as follows. In \cref{sec:uldm} we review the description of the spin-1 and spin-2 ultra-light dark matter fields. We derive the signal morphology and confirm it through simulations in \cref{sec:signalmorph}; \cref{sec:method} describes the new \bsdxcorr implementation and its sensitivity. In \cref{sec:impact} we discuss the impact of using the proposed optimal search integration time on existing pipelines and demonstrate the projected improvement in O4a sensitivity limits. We contextualise our results and conclude in \cref{sec:end}.

%%%%%%%%%%%%%%%%%%%%%%%%%%%%%%%%%%%%%%%%%%%%
\section{Ultra-light dark matter fields}
\label{sec:uldm}
%%%%%%%%%%%%%%%%%%%%%%%%%%%%%%%%%%%%%%%%%%%%

We begin by reviewing the description of the coherent oscillation of the spin-1 and spin-2 ultra-light dark matter, as is relevant for their effect on the interferometers. First of all, because of the extremely large occupation number of ultra-light dark matter, the field can be described as a superposition of classical waves as
\begin{align}\label{eq:uldm_def}
    \field(t,\vx)=\sum_{a=1}^N \field_0^a \cos\left(\omega_a t-\vk_a\cdot\vx+\gamma_a\right),
\end{align}
where $\field$ is taken to represent the components of a bosonic field of any integer spin. For each wave indexed by $a$, $\field_0^a$ represents its amplitude and it encodes its spin-dependent polarisation structure (e.g.\ it is a scalar for spin~0, a vector for spin~1 and a tensor for spin~2, and so on), $\omega_a$ is the angular frequency, $\vk_a$ is the wave number, and $\gamma_a$ is a constant phase. Each wave has a different velocity $v_a$ distributed according to the following Maxwell-Boltzmann (MB) distribution around $v_0$:
\begin{align}\label{eq:MBPDF}
    {\rm PDF}(v)=\frac{4v^2}{\sqrt{\pi}v_0^3}\,{\rm e}^{-v^2/v_0^2}.
\end{align}
Because of the non-relativistic character of the particle-wave, the angular frequency $\omega_a$ of each wave is related to the ultra-light dark matter mass $m$ and the velocity $v_a$ via
\begin{align}\label{eq:omegaandmrelation}
    \omega_a = m\left(1+\frac12v_a^2+\mathcal{O}\left(v_a^4\right)\right).
\end{align}
Because the angular frequency of each wave is slightly different, the superposition of all waves $\field$ will, therefore, not be monochromatic, but exhibit a characteristic frequency broadening scaling with $v^2$, see \cref{sec:distinguish}.

In the following sections we focus on two models: the spin~1 model, also called dark photon or dark vector, and the spin~2 model, also called dark graviton or dark tensor. In the first case we write the field components as
\begin{align}\label{eq:vec_def}
    \field \doteq A_i, \quad& \field_0^a \doteq A_0^a \varepsilon_i^a,
\end{align}
where $A_i$ are the three spatial components of the vector field, $A_0^a$ denotes each wave's amplitude, and $\varepsilon_i^a$ is each wave's polarisation. For the tensor case we write
\begin{align}\label{eq:ten_def}
    \field \doteq M_{ij}, \quad& \field_0^a \doteq M_0^a \varepsilon_{ij}^a,
\end{align}
where $M_{ij}$ contains the five physical components of the tensor field, $M_0^a$ denotes each wave's amplitude, and $\varepsilon_{ij}^a$ is each wave's polarisation.

%%%%%%%%%%%%%%%%%%%%%%%%%%%%%%%%%%%%%%%%%%%%
\subsection{Spin-1 ultra-light dark matter}
\label{sec:spin1uldm}

Let us consider the spin-1 ultra-light dark matter case of \cref{eq:vec_def}; rewriting \cref{eq:uldm_def} explicitly for this case, we find
\begin{align}\label{eq:spin1field}
    A_i(t,\vx) = \sum_{a=1}^N A_0^a \varepsilon_i^a \cos\left(\omega_a t-\vk_a\cdot\vx+\gamma_a\right),
\end{align}
where, in the non-relativistic limit, the field time component $A_0(t,\vx)$ (not to be confused with the wave amplitudes $A_0^a$) is $v_0$-suppressed with respect to the spatial components $A_i(t,\vx)$ and we can safely ignore it. The dark matter basis $(\boldsymbol{p},\boldsymbol{q},\boldsymbol{r})$ is related to the detector frame via 
\begin{align}\label{eq:pqr}
\nonumber    \boldsymbol{r}&=(\sin\theta \cos\phi, \sin\theta \sin\phi, \cos\theta),\\
\nonumber    \boldsymbol{p}&=(\cos\theta \cos\phi, \cos\theta \sin\phi, -\sin\theta),\\
             \boldsymbol{q}&=(-\sin\phi, \cos\phi, 0),
\end{align}
where $(\theta,\phi)$ are the spherical angles of the detector frame. In this basis, the three polarisations of the vector field, \(\{\varepsilon_p,\varepsilon_q,\varepsilon_r\}\), can be written as
\begin{align}\label{eq:s1plo_def}
    \varepsilon^a_i \doteq \sum_\kappa \varepsilon_\kappa {\cal Y}^\kappa_i,
\end{align}
where \(\sum_\kappa \varepsilon_\kappa^2=1\) and \({\cal Y}^p_i = p_i\), \({\cal Y}^q_i = q_i\) and \({\cal Y}^r_i = r_i\).

The interaction between the spin-1 ultra-light dark matter and standard matter is given by~\cite{Manita:2023mnc}
\begin{align}\label{eq:Lintspin1}
    \mathcal{L}_{\rm int}=\epsilon_D e A_\mu j^\mu_D,
\end{align}
where $j_D^\mu$ denotes the current of the Standard Model particles for either baryon, i.e.\ $D=B$, or baryon-minus-lepton, $D=B-L$, symmetry. This coupling causes the mirrors of gravitational wave detectors to experience an acceleration described by the force
\begin{align}\label{eq:forceonmirrorspin1}
    F_i=-\epsilon_D e Q_D \dot{A_i}(t,\vx),
\end{align}
where $\epsilon_D$ is the coupling constant, $e$ is the electric charge, and $Q_D$ is the $U(1)$ charge of the mirror.  

%%%%%%%%%%%%%%%%%%%%%%%%%%%%%%%%%%%%%%%%%%%%
\subsection{Spin-2 ultra-light dark matter}
\label{sec:spin2uldm}

In the case of spin-2 ultra-light dark matter instead, upon using \cref{eq:ten_def}, we find
\begin{align}\label{eq:spin2field}
    M_{ij}(t,\vx)=\sum_{a=1}^N M_0^a\varepsilon_{ij}^a \cos\left(\omega_a t-\vk_a\cdot\vx+\gamma_a\right).
\end{align}
The polarisation tensor is described as~\cite{Armaleo:2020efr}
\begin{align}\label{eq:s2pol_def}
    \varepsilon_{ij}^a \doteq \sum_\kappa \varepsilon_\kappa \cY^\kappa_{ij},
\end{align}
where $\kappa$ labels the five polarisation amplitudes, $\{\varepsilon_+, \varepsilon_\times,\varepsilon_L,\varepsilon_R,\varepsilon_S\}$, which satisfy $\sum_\kappa \varepsilon_\kappa^2=1$, and
\begin{alignat}{2}
    \cY^\times_{ij}&\doteq \frac{1}{\sqrt{2}}(p_i q_j+q_i p_j),&\quad
    \cY^+_{ij}&\doteq \frac{1}{\sqrt{2}}(p_i p_j-q_i q_j),\nonumber\\
    \cY^L_{ij}&\doteq \frac{1}{\sqrt{2}}(q_i r_j+r_i q_j),&\quad
    \cY^R_{ij}&\doteq \frac{1}{\sqrt{2}}(p_i r_j+r_i p_j),\nonumber\\
    \cY^S_{ij}&\doteq \frac{1}{\sqrt{6}}(3r_i r_j-\delta_{ij}).\label{eq:Ybasis}
\end{alignat}
Notice that, similarly to the vector case, for a plane wave in the non-relativistic limit, the linearised equations of motion in vacuum enforce $M_{00} \propto v_0 M_{0i} \propto v_0^2 M_{ij}$ and we can therefore safely neglect the subdominant time components of the spin-2 field.

The spin-2 field couples to the Standard Model via~\cite{Armaleo:2020efr}
\begin{align}\label{eq:Lintspin2}
    \mathcal{L}_{\rm int} = \frac{\alpha}{M_{\rm Pl}} \sqrt{-g} M_{ij} T^{ij}_{\rm SM},
\end{align}
where $T^{ij}_{\rm SM}$ is the energy-momentum tensor of Standard Model fields. At linear order in $\alpha$, we can absorb this interaction into the metric definition, resulting in a description that is equivalent to a metric perturbation $h_{ij}=2\alpha M_{ij}/M_{\rm Pl}$. Therefore, the spin-2 field causes a strain in the detectors similar to a usual tensor perturbation produced by a gravitational wave. 

%%%%%%%%%%%%%%%%%%%%%%%%%%%%%%%%%%%%%%%%%%%%
\section{Signal morphology}
\label{sec:signalmorph}
%%%%%%%%%%%%%%%%%%%%%%%%%%%%%%%%%%%%%%%%%%%%

%%%%%%%%%%%%%%%%%%%%%%%%%%%%%%%%%%%%%%%%%%%%
\subsection{Detector response}
\label{sec:detectorresponse}

In this section we describe the detector responses for the spin-1 and spin-2 ultra-light dark matter signals in an L-shaped gravitational wave detector, following the treatment of~\cite{Manita:2023mnc}. For a more detailed derivation of these results see \cref{app:strains}.

%%%%%%%%%%
\paragraph{Response to the spin-1 signal.}
Given the force on the detector mirrors, \cref{eq:forceonmirrorspin1}, we readily obtain the acceleration projected in the $\hat{\vx}$ direction, chosen as the direction of one of the arms
\begin{align}\label{eq:eomspin1}
    \frac{d^2x}{dt^2}=-\epsilon_D e \frac{Q_D}{M} \hat{x}_i \dot{A}^i,
\end{align}
where $M$ is the mirror mass. The differential strain that arises from this acceleration is
\begin{align}\label{eq:strainspin1ts}
    h \doteq &\, h_{\rm t}+h_{\rm s} \nonumber\\
    =&\, 2\frac{\epsilon_D e \, Q_D}{ML} \sum_{a=1}^N \sin^2\left(\frac{\omega_a L}{2}\right) \frac{1}{\omega_a} A_0^a \varepsilon_i^a D^{i}\sin[\omega_a(t-L)+\gamma_a] \nonumber\\
    & -2 \frac{\epsilon_D e \, Q_D}{M}\sum_{a=1}^N v_a A_0^a \varepsilon_i^a \hat{k}_j D^{ij} \cos[\omega_a(t-L)+\gamma_a],
\end{align}
where the two terms, \(h_\mathrm{t}\) and \(h_\mathrm{s}\) are physically distinct contributions that refer to the displacement of mirrors during the round-trip of the laser, known as the finite-time travelling effect, and the difference in spatial displacement between the input and end mirrors, respectively. The proper displacement strain induced by spin-1 ultra-light dark matter, \(h_\mathrm{s}\), is proportional to the momentum of the ultra-light dark matter, because the mirrors acted upon by the force \cref{eq:eomspin1} move in the same direction; the first-order differential effect is therefore due to the gradient of the field, unlike in the spin-2 case (or in the case of a gravitational wave), where the differential strain appears already at zeroth order, see below. The response functions for the two terms are defined as
\begin{align}\label{eq:responsespin1}
    D_i \doteq (\hat{x}_i -\hat{y}_i),
\end{align}
in the `time' contribution and, for the `space' contribution, as
\begin{align}\label{eq:responsespin2}
    D_{ij} \doteq \frac{1}{2}(\hat{x}_i\hat{x}_j -\hat{y}_i\hat{y}_j).
\end{align}
Notice that in practice the detector experiences only the combined overall strain of \cref{eq:strainspin1ts}, but because of the different response structure, the relative weight of the two terms depends on the detector geometry.

The strain we just derived in \cref{eq:strainspin1ts} depends explicitly on time, the unknown phase and exact velocity of each ultra-light dark matter wave, its orientation with respect to the detector frame and polarisation. Hence, in order to estimate the strain we must average the effect of the distribution of velocities, as well as over time and angles. We refer the reader to \cref{app:strains} and here we only report the final result, which reads
\begin{align}\label{eq:avrstrainspin1}
    \sqrt{\left<h^2\right>} =&\, \frac{2\epsilon_D e Q_D \sqrt{ \rho_{\rm DM}}}{\sqrt{3}\,m M} \left[\frac{2}{(mL)^2}\sin^4\left(\frac{mL}{2}\right)+\frac{3}{10}v_0^2\right]^{\frac{1}{2}}
\end{align}
where $\left<.\right>$ denotes the averages over time, velocity, polarisation and direction. In terms of the time and space contributions to the strain this gives
\begin{align}
    \sqrt{\left<h_{\rm t}^2\right>} &\simeq 1.32 \times 10^{-25} \left(\frac{\epsilon_D}{10^{-22}}\right), \nonumber\\
    \sqrt{\left<h_{\rm s}^2\right>} &\simeq 1.87 \times 10^{-26} \left(\frac{\epsilon_D}{10^{-22}}\right)\left(\frac{100{\rm Hz}}{f}\right),
\end{align}
where $f=mc^2/(2\pi\hbar)$.\footnote{Note that the conversion from the mass $m$ in eV to the frequency $f$ in Hz is made with the conversion factor in SI units, namely $1{\rm eV} = 2.42\times 10^{14} {\rm Hz}$.}

%%%%%%%%%%
\paragraph{Response to the spin-2 signal.}
Given the coupling from \cref{eq:Lintspin2} between the spin-2 field and Standard Model particles, the displacements of the detector arms is obtained from the geodesic deviation equation
\begin{align}\label{eq:geodesiceq}
    \frac{d^2 x^j}{dt^2}=-R^j_{\;0k0} x^k,
\end{align}
where $R_{j0k0}$ is the Riemann tensor formed from the metric 
\begin{align}
    g_{\mu\nu}=\eta_{\mu\nu}+2\alpha \frac{M_{\mu\nu}}{M_{\rm Pl}},
\end{align}
and \(\eta_{\mu\nu}\) is the Minkowski metric. From $\partial_\mu M^{\mu\nu}=0$, obtained from the equation of motion in vacuum, we have $M_{0\mu}\ll M_{ij}$. Neglecting the $0\mu$ components we find
\begin{align}
    R_{j0k0}\simeq -\frac{\alpha}{M_{\rm Pl}}\ddot{M}_{jk}. 
\end{align}
Substituting in \cref{eq:geodesiceq} we have
\begin{align}\label{eq:geodesiceq2}
    \frac{d^2x}{dt^2}=\frac{\alpha x}{M_{\rm Pl}}\ddot{M}_{jk} \hat{x}^j \hat{x}^k.
\end{align}

In the spin-2 case the strain can also be separated into two physically distinct effects, the finite-time travelling effect and the standard, gravitational-wave-like, spatial strain:
\begin{align}\label{eq:strainspin2ts}
    h \doteq&\, h_{\rm t}+h_{\rm s}\nonumber\\
    = &\, \frac{2\alpha}{M_{\rm Pl}}D^{ij}\sum_{a=1}^N \sin^2\left(\frac{\omega_a L}{2}\right) M_0^a\varepsilon_{ij}^a \cos\left(\omega_a (t-L)+\gamma_a\right) \nonumber\\
    & -\frac{2\alpha}{M_{\rm Pl}}D^{ij}\sum_{a=1}^N M_0^a\varepsilon_{ij}^a \cos\left(\omega_a (t-L)+\gamma_a\right).
\end{align}
Just as we did in the previous case, in order to numerically estimate the strain we perform an average over the wave's velocities, time, polarisation and direction to obtain
\begin{align}
\label{eq:avrstrainspin2}
    \sqrt{\left<h^2\right>}&=\frac{\alpha \sqrt{2\rho_{\rm DM}}}{\sqrt{5} \,m M_{\rm Pl}} \left[1+\sin^4\left(\frac{m L}{2}\right)\right]^{\frac{1}{2}} \nonumber\\
    &\simeq 1.12 \times 10^{-25} \left(\frac{\alpha}{10^{-7}}\right) \left(\frac{100 {\rm Hz}}{f}\right).
\end{align}

%%%%%%%%%%%%%%%%%%%%%%%%%%%%%%%%%%%%%%%%%%%%
\subsection{Sidereal modulation}\label{sec:siderealmodulation}
Ground-based detectors experience a sidereal motion due to Earth's rotation, causing a modulation of the signal that depends on how a specific field couples to the detector. In this section we derive the sidereal modulation for both spin-1 and spin-2 ultra-light dark matter and investigate whether and how it is possible to distinguish these particles from each other and from continuous gravitational waves of astrophysical origin.

%%%%%%%%%%
\paragraph{Sidereal modulation of the spin-1 signal.}
As we have shown in \cref{sec:detectorresponse}, the coupling between spin-1 ultra-light dark matter and the gravitational wave detectors can be written as
\begin{align}\label{eq:strainspin1}
    h(t) \simeq h_{0,\mathrm{t}}\sin{(m t+\gamma)}\Delta \varepsilon_{\mathrm{t}} + h_{0,\mathrm{s}}\cos{(m t+\gamma)}\Delta \varepsilon_{\mathrm{s}},
\end{align}
where
\begin{align}\label{eq:h0def}
    h_{0,\mathrm{t}} \doteq 2\epsilon_D e \frac{Q_D}{M}\frac{1}{\omega L}\sin^2{\frac{\omega L}{2}}A_0, \quad h_{0,\mathrm{s}} \doteq - 2\epsilon_D e \frac{Q_D}{M} v A_0.
\end{align}
The contractions
\begin{align}\label{eq:deltaepsilonsspin1}
    \Delta\varepsilon_{\mathrm{t}} & \doteq \varepsilon_{i}(\hat{x}^i-\hat{y}^i), \quad \Delta\varepsilon_{\mathrm{s}} \doteq \frac12 \varepsilon_{i} \hat{k}_j (\hat{x}^i\hat{x}^j-\hat{y}^i\hat{y}^j),
\end{align}
describe the geometry of the detector response to ultra-light dark matter. The $mL=\omega L$ constant terms in the oscillating functions are absorbed in the phase $\gamma$, as they simply represent a time shift. Notice that, as it will become clear below, for the purpose of the discussion in this section we can assume that all waves have the same velocity, amplitude, phase and polarisations, so we drop the $a$ index. Lastly, without loss of generality, we can align $\hat{\vk} \parallel \hat{\boldsymbol{r}}$.

The directions of the detector's arms $\hat{x}^i$ and $\hat{y}^i$ are a function of time due to the sidereal (rotational) motion, for which the rotation angle $\theta_\mathrm{sid}$ depends on a phase that defines Earth's position at $t=0$ and on the rotational angular velocity of Earth~\cite{Jaranowski:1998qm}, such that
\begin{align}
    \theta_\mathrm{sid}=\phi_r+\Omega_r t,
\end{align}
where $\Omega_r=2\pi/T_\mathrm{sid}\simeq 7.29\times 10^ {-5}\,{\rm Hz}$ and $\phi_r$ is a phase that relates the global sidereal time with the local sidereal time of the detector. The motion of the detector's arms is described by the rotation matrix
\begin{align}\label{eq:siderealrotation}
    R_\mathrm{sid} = \begin{pmatrix}
    \sin\lambda \cos\theta_\mathrm{sid}& -\sin\theta_\mathrm{sid} & \cos\lambda \cos\theta_\mathrm{sid} \\
    \sin\lambda\sin\theta_\mathrm{sid} & \cos\theta_\mathrm{sid} & \cos\lambda \sin\theta_\mathrm{sid} \\
    -\cos\lambda & 0 & \sin\lambda  \\
\end{pmatrix},
\end{align}
acting on the initial arm unit vectors. Here $\lambda$ is the latitude of the detector, namely $\lambda_H=46.45\degree$ for LIGO-Hanford and $\lambda_L=30.56\degree$ for LIGO-Livingston. In order to correctly account for the relative orientation of the LVK detectors, we apply a further detector-dependent rotation
\begin{align}\label{eq:siderealrotationM3}
    R_\mathrm{det} = \begin{pmatrix}
    -\sin\xi& -\cos\xi & 0 \\
    \cos\xi & -\sin\xi & 0 \\
    0 & 0 & 1  \\
\end{pmatrix},
\end{align}
to each detector for which the $\hat{x}^i_0$ and $\hat{y}^i_0$ arms orientations are chosen as $\hat{\vx}$ and $\hat{\vy}$ respectively.\footnote{In the language of~\cite{Jaranowski:1998qm} the rotation \(R(\theta_\mathrm{sid})\) in \cref{eq:siderealrotation} corresponds to \(M_2^T\), whereas \cref{eq:siderealrotationM3} is the \(M_3^T\) rotation matrix.} Here we defined \(\xi\doteq\psi+\zeta/2\), where \(\psi\)
determines the orientation of the detector’s arms with respect to local geographical directions (\(\psi = 171.8^\circ\) and \(
243.0^\circ\) for LIGO-Hanford and LIGO-Livingston, respectively) and \(\zeta\) is the angle between the interferometer arms (in LVK \(\zeta=\pi/2\)). The detector's arms are therefore described by \(\hat{\vx}(t)= R_\mathrm{sid} \, R_\mathrm{det} \, \hat{\vx}_0\) and \(\hat{\vy}(t)= R_\mathrm{sid} \, R_\mathrm{det} \, \hat{\vy}_0\).
Using these expressions we find that the $\Delta\varepsilon_{\rm t}$ and $\Delta\varepsilon_{\rm s}$ terms become
\begin{align}
    \Delta\varepsilon_{\mathrm{t}} &= \Delta\varepsilon^{\mathrm{t}}_{-1} \sin{\theta_\mathrm{sid}} + \Delta\varepsilon^{\mathrm{t}}_{0} + \Delta\varepsilon^{\mathrm{t}}_{1} \cos{\theta_\mathrm{sid}}, \label{eq:dep_s1t}\\
    \Delta\varepsilon_{\mathrm{s}} &= \Delta\varepsilon^{\mathrm{s}}_{-2} \sin{2\theta_\mathrm{sid}} + \Delta\varepsilon^{\mathrm{s}}_{-1} \sin{\theta_\mathrm{sid}} + \Delta\varepsilon^{\mathrm{s}}_{0} + \Delta\varepsilon^{\mathrm{s}}_{1} \cos{\theta_\mathrm{sid}} + \Delta\varepsilon^{\mathrm{s}}_{2} \cos{2\theta_\mathrm{sid}}, \label{eq:dep_s1s}
\end{align}
where the coefficients $\Delta\varepsilon^{\rm t,s}_q$ are functions of the angles \(\theta\), \(\phi\), \(\lambda\) and \(\xi\), and of the three polarisation amplitudes $\varepsilon_p,~\varepsilon_q$, and~$\varepsilon_r$, see \cref{app:peaks}.

Recasting the strain \cref{eq:strainspin1} in the frequency domain, after some algebra we find that the signal will present multiple peaks located, for a wave with \(v=0\), at $\omega=m\pm q\Omega_r$: for the \(h_{0,\mathrm{t}}\) contribution we have \(q\in\{0,1\}\), whereas for the \(h_{0,\mathrm{s}}\) term we have \(q\in\{0,1,2\}\) -- the non-zero velocity of each wave will move the peaks to the right according to \cref{eq:omegaandmrelation}. This harmonic structure is a direct analogue of the peaks of continuous GWs~\cite{Jaranowski:1998qm}. The number of peaks, or harmonics, depends on the index structure of the ultra-light dark matter field and the detector response: up to three for a spin-1 field and up to five for a spin-2 field at zeroth order in gradients, increasing by two with each consecutive term in the gradient expansion -- notice that the \(h_{0,\mathrm{t}}\) is a zeroth-order term whereas \(h_{0,\mathrm{s}}\) is a first-order one. The total separation of the three peaks of the \(h_{0,\mathrm{t}}\) term is $\Delta f_{\rm sid,\,t}^{\rm spin-1}\simeq 2.32\times10^{-5}{\rm~Hz}$, whereas for the five peaks of the \(h_{0,\mathrm{s}}\) term the total separation is $\Delta f_{\rm sid,\,s}^{\rm spin-1} \simeq 4.64\times10^{-5}{\rm~Hz}$. 

The relative amplitudes of the peaks are sensitive to the location and orientation of each detector; nevertheless, we can infer some general properties. First, for frequencies \(f\gtrsim9\,\mathrm{~Hz}\) the \(h_{0,\mathrm{t}}\) is numerically dominant; we therefore expect that the dominant signal contribution will come from (some of) the three \(q\in\{0,1\}\) peaks, with additional, subdominant signal in the \(q=2\) peaks. Notice that this does not mean that the three (or five) peaks will be equally populated, in fact we will see below that for LIGO-Hanford the \(q=0\) central peak is suppressed with respect to the \(q=1\) peaks, whereas for LIGO-Livingston that is not the case; looking at the coefficients reported in \cref{app:peaks} we see that this is because the orientation of the arms for LIGO-Hanford sits very close to \(\xi=225^\circ\), for which \(\Delta\varepsilon_0^\mathrm{t} \approx \Delta\varepsilon_0^\mathrm{s} \approx 0\). At lower frequencies the \(h_{0,\mathrm{s}}\) term dominates and the signal can in principle populate all five peaks, although, once again, their relative ratio depends on the specific location and orientation of the detector. 

%%%%%%%%%%
\paragraph{Sidereal modulation of the spin-2 signal.}
In \cref{sec:detectorresponse} we obtained that the spin-2 strain is given by
\begin{align}\label{eq:h}
    h(t)=h_0\cos{(m t+\gamma)}\Delta \varepsilon,
\end{align}
where $h_0$ is given by 
\begin{align}\label{eq:h0def2}
    h_0\doteq \frac{4\alpha \sqrt{\rho_{\rm DM}}}{\sqrt{2}M_{\rm Pl}m}\left[-1+\sin^2\left(\frac{\omega L}{2}\right)\right],
\end{align}
and 
\begin{align}
    \Delta\varepsilon\doteq \frac12 \varepsilon_{ij}(\hat{x}^i \hat{x}^j-\hat{y}^i \hat{y}^j)
\end{align}
describes the contraction between the polarisation tensor $\varepsilon_{ij}$ of the spin-2 field and the detector response of \cref{eq:responsespin2}. Applying the same procedure as in the case of spin~1 we find that the sidereal modulation of the spin-2 ultra-light dark matter signal takes the form
\begin{align}\label{eq:dep_s2}
    \Delta\varepsilon &= \Delta\varepsilon_{-2} \sin{2\theta_\mathrm{sid}} + \Delta\varepsilon_{-1} \sin{\theta_\mathrm{sid}} + \Delta\varepsilon_{0} + \Delta\varepsilon_{1} \cos{\theta_\mathrm{sid}} + \Delta\varepsilon_{2} \cos{2\theta_\mathrm{sid}}
\end{align}
where, once again, $\Delta\varepsilon_q$ are functions of the four angles \(\theta\), \(\phi\), \(\lambda\) and \(\xi\), and of the five polarisation amplitudes $\varepsilon_+,~\varepsilon_\times,~\varepsilon_L,~\varepsilon_R$, and~$\varepsilon_S$, see \cref{app:peaks}. For $\theta_\mathrm{sid}=0$, $\lambda=\pi/2$, and $\xi=-\pi/2$, we recover the result obtained for time-independent arms' orientations~\cite{Armaleo:2020efr}. Moving on to the expression for the strain in the frequency domain, we find that the signal has five peaks, as expected from the spin-2 structure, located at $\omega=m\pm2\Omega_r$ with \(q\in\{0,1,2\}\), amounting to a total separation of peaks of $\Delta f_{\rm sid}^{\rm spin-2}\simeq 4.64\times10^{-5}{~\rm Hz}$. Also, similarly to the spin-1 case, we see that for LIGO-Hanford the central peak is suppressed.

%%%%%%%%%%%%%%%%%%%%%%%%%%%%%%%%%%%%%%%%%%%%
\subsection{Differentiating between signals}
\label{sec:distinguish}

Let us now investigate whether differences in the morphology of the spin-1 and spin-2 signals in the frequency domain can be used to distinguish the nature of the signal in an eventual detection. We have seen that the morphology of the signal is, for most of the LVK frequency range, different for the spin-1 and spin-2 models. In the former, besides the central harmonic at \(\omega=m\), four peaks appear due to the sidereal modulation, two coming from the contribution ($\omega=m\pm\Omega_r$), namely the finite-time travelling effect $\delta L_{\rm t}$, which is dominant at \(f\gtrsim9\,\mathrm{Hz}\), and a further two coming from the subdominant part of the response ($\omega=m\pm2\Omega_r$), namely the spatial displacement $\delta L_{\rm s}$.\footnote{Strictly speaking, there are three plus five peaks but the central three overlap, so we speak of three plus two here.} This happens because, in the spin-1 case, the time and space contributions lead to different behaviours of the detector response, which is proportional to one or two powers, respectively, of the unit vectors defining the detector arms. For the spin-2 signal both response types have five peaks (the central one with \(\omega=m\) and four more at $\omega=m\pm\Omega_r$ and $\omega=m\pm2\Omega_r$) because both $\delta L_{\rm t}$ and $\delta L_{\rm s}$ have the same response geometry.

Besides the appearance of multiple peaks due to the sidereal motion of Earth-bound detectors, the frequency-domain signal will experience a broadening reflecting the fact  that ultra-light dark matter waves have different velocities distributed according to the MB distribution of \cref{eq:MBPDF}, and are therefore not perfectly monochromatic. The frequency spread is given by
\begin{align}\label{eq:MBestimate}
    \Delta f_v=\frac{1}{2}\left(\frac{v_{\rm esc}}{c}\right)^2 f\simeq 1.58 \times 10^{-6} f,
\end{align}
where we have conservatively chosen the dark matter escape velocity $v_{\rm esc}=533~{\rm km/s}$ ($1.78 \times 10^{-3}c$) as the largest velocity the dark matter waves could possess. Therefore, the multiple peaks are only clearly visible for frequencies $f$ satisfying
\begin{align}
    \Delta f_{\rm sid}>\Delta f_v,
\end{align}
where $\Delta f_{\rm sid}=4.64\times 10^ {-5}$~Hz applies to both spin-1 and spin-2 signals, resulting in $f\lesssim 30\,\mathrm{Hz}$.\footnote{In this section we are referring to the largest possible separation. Note that for the spin-1 case, this is given by the space contribution to the strain $\Delta f_{\rm sid,s}^{\rm spin-1}$.} If we adopt the more strict criterion that the peaks do not overlap, then \(\Delta f_v < 1/T_\mathrm{sid}\), which gives $f\lesssim 7\,\mathrm{Hz}$. However, since the sidereal peaks are an intrinsic feature of the signal, they contribute to its frequency spread across the entire frequency range. As a result, the total broadening is the combined effect of the MB spread \(\Delta f_v\) and the additional spread due to the peak separation \(\Delta f_\mathrm{sid}\).

\begin{figure}
\begin{center}
\includegraphics[width=0.7\linewidth]{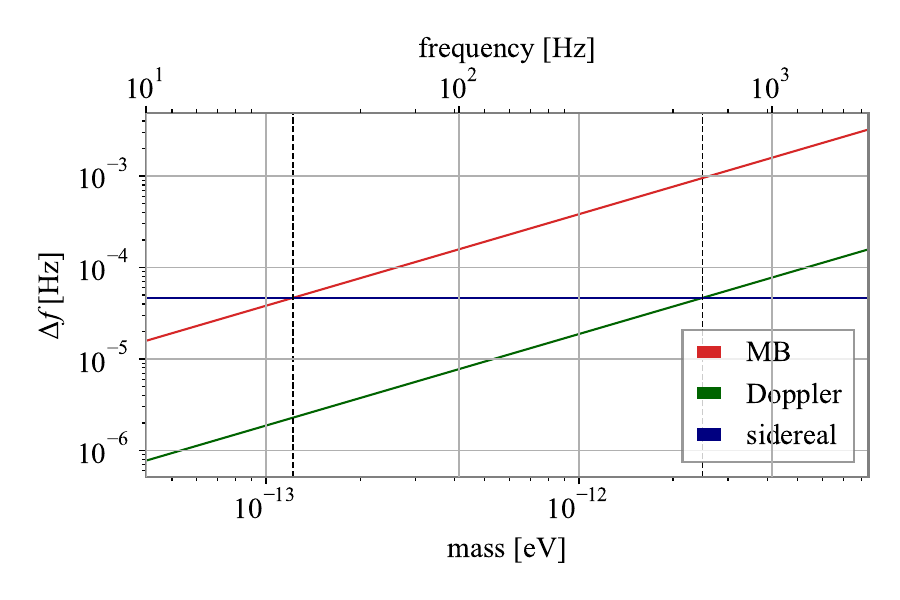}
\end{center}
\caption{MB ($\Delta f_v$), Doppler ($\Delta f_{\rm Doppler}$) and sidereal ($\Delta f_{\rm sid}$) frequency spreads suffered by the ultra-light dark matter signals as a function of frequency (or dark matter mass). The vertical lines correspond to $\Delta f_{\rm sid}=\Delta f_v~(\sim 30~{\rm Hz})$ and $\Delta f_{\rm sid}=\Delta f_{\rm Doppler}~(\sim 600~{\rm Hz})$, respectively.
} \label{fig:modulations}
\end{figure}

In addition to the sidereal and MB frequency spreads, the signal is also affected by the Doppler modulation, which is caused by the Earth's movement with respect to the ultra-light dark matter directions $\hat{k}_i$. Such modulation is computed as $\Delta f_{\rm Doppler}\simeq k v_E/(2\pi)$,  where $v_E$ includes the orbital and rotational velocities of Earth, and has been shown to be subdominant with respect to the MB spread in \cref{eq:MBestimate}~\cite{Miller:2020vsl}. When compared to the sidereal modulation, the Doppler effect is also subdominant for $f\lesssim 600~$Hz, while for higher frequencies both the Doppler and sidereal modulations are much smaller than the MB frequency spread $\Delta f_v$. The hierarchy between the modulations is shown in \cref{fig:modulations}. For this reason, we neglect the Doppler modulation from now on.

\begin{figure}
\begin{center}
\includegraphics[width=0.9\linewidth]{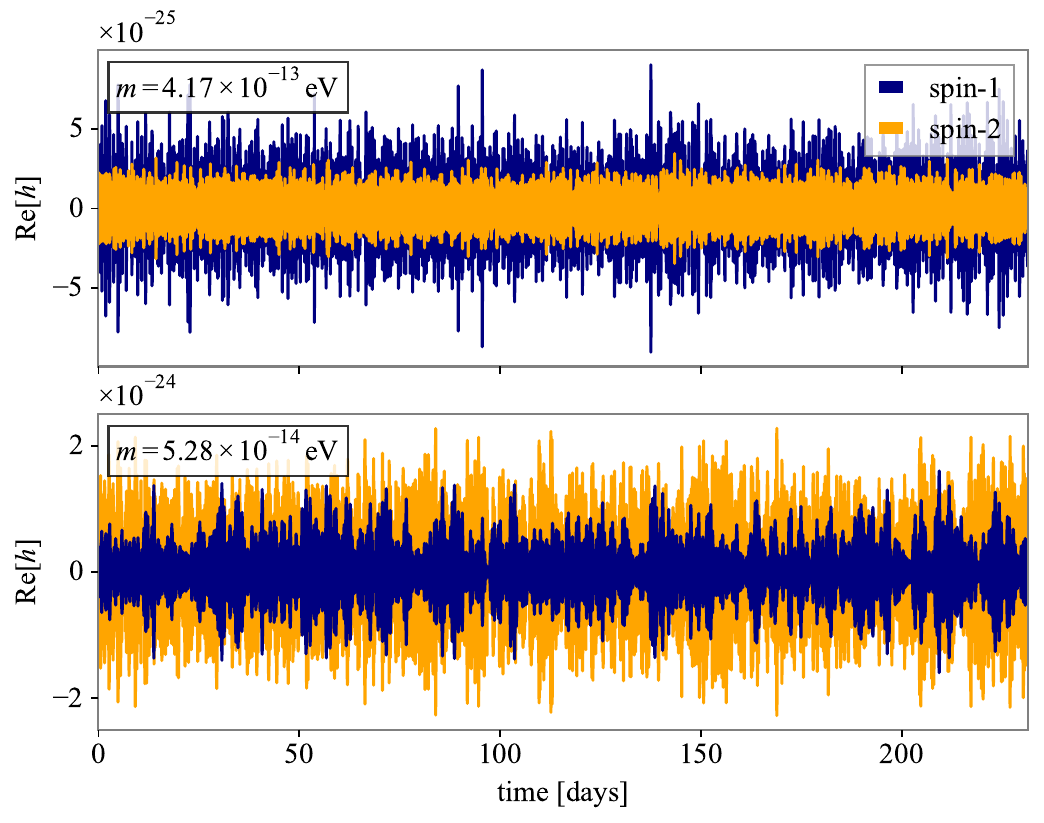}
\end{center}
\caption{Noiseless strain in Hanford obtained from the simulation of 1000 spin-1 and spin-2 ultra-light dark matter waves contracted with the corresponding detector responses, including the sidereal modulation, with $\epsilon_D=8.52 \times 10^{-23}$ for the spin-1 signal and $\alpha=10^{-7}$ for the spin-2 signal. The masses correspond to (top) $m=4.17\times 10^{-13}$eV and (bottom) $m=5.28\times 10^{-14}$eV for both spins. The strains in Livingston are similar to the ones in Hanford, with the difference being the detector orientation parameters used in the sidereal modulation. \label{fig:strain}
}
\end{figure}

In order to visualise all these effects, we simulate the ultra-light dark matter signal (of spin~1 or spin~2) first in zero noise. A more realistic simulation, adding the signal to real LVK data, is also described in \cref{sec:method}. All the simulations have been performed using the  BSD modules of the Matlab-based Snag toolbox~\cite{Piccinni:2018akm,Frasca:2000ni}. The simulated signals are stored in the same format as the BSD files, which contain appropriately sampled and band-limited time series in the so-called reduced-analytic signal format~\cite{Piccinni:2018akm}.
In practice, we simulate the dark matter waves in \cref{eq:uldm_def} by superposing $1000$~waves  spanning $236.98~$days of data and $1~$Hz frequency range (see \cref{fig:strain}).\footnote{In our halo the occupation number of the ultra-light dark matter field is much higher than 1000. However, 1000~simulated waves are enough to correctly sample the signal amplitude and frequency dispersion.}

For spin-1 waves the polarisation tensor of each wave 
\begin{align}
    \vep_a \doteq (\varepsilon_p,\varepsilon_q,\varepsilon_r) = (\sin\vartheta\cos\varphi,\sin\vartheta\sin\varphi,\cos\vartheta)
\end{align}
is randomised via a uniform sampling for $\varphi$ and an isotropic polar sampling for $\vartheta$ in order to uniformly cover the sphere. For the spin-2 case the polarisation amplitudes $\varepsilon_+,~\varepsilon_\times,~\varepsilon_L,~\varepsilon_R$, and $\varepsilon_S$ of each wave are randomised in such a way that they isotropically sample the 4-sphere.

In both the spin-1 and spin-2 cases, the velocities are distributed according to \cref{eq:MBPDF} via rejection sampling. Lastly, for the field \(\field\) to correctly account for the entirety of the cosmological dark matter, the total amplitude of the superposition of waves is required to lead to the expected dark matter density in the solar region, $\rhoDM = 0.40~\GeV/\cm^3$, via
\begin{align}\label{eq:ampnorm}
    \field_\mathrm{norm} = \frac{\field}{\sqrt{\mathcal{I}}}, \quad \mathrm{with~} \mathcal{I} &= \frac{1}{\rhoDM}\frac{1}{T} \int\!\de t \, m^2 |\field|^2,
\end{align}
where $T$ represents the total duration of the simulated field~\cite{Guo:2019ker}. In what follows we will omit the `$\mathrm{norm}$' index for simplicity, with the understanding that this normalisation is always implemented.

\begin{figure}
\begin{center}
\includegraphics[width=0.9\linewidth]{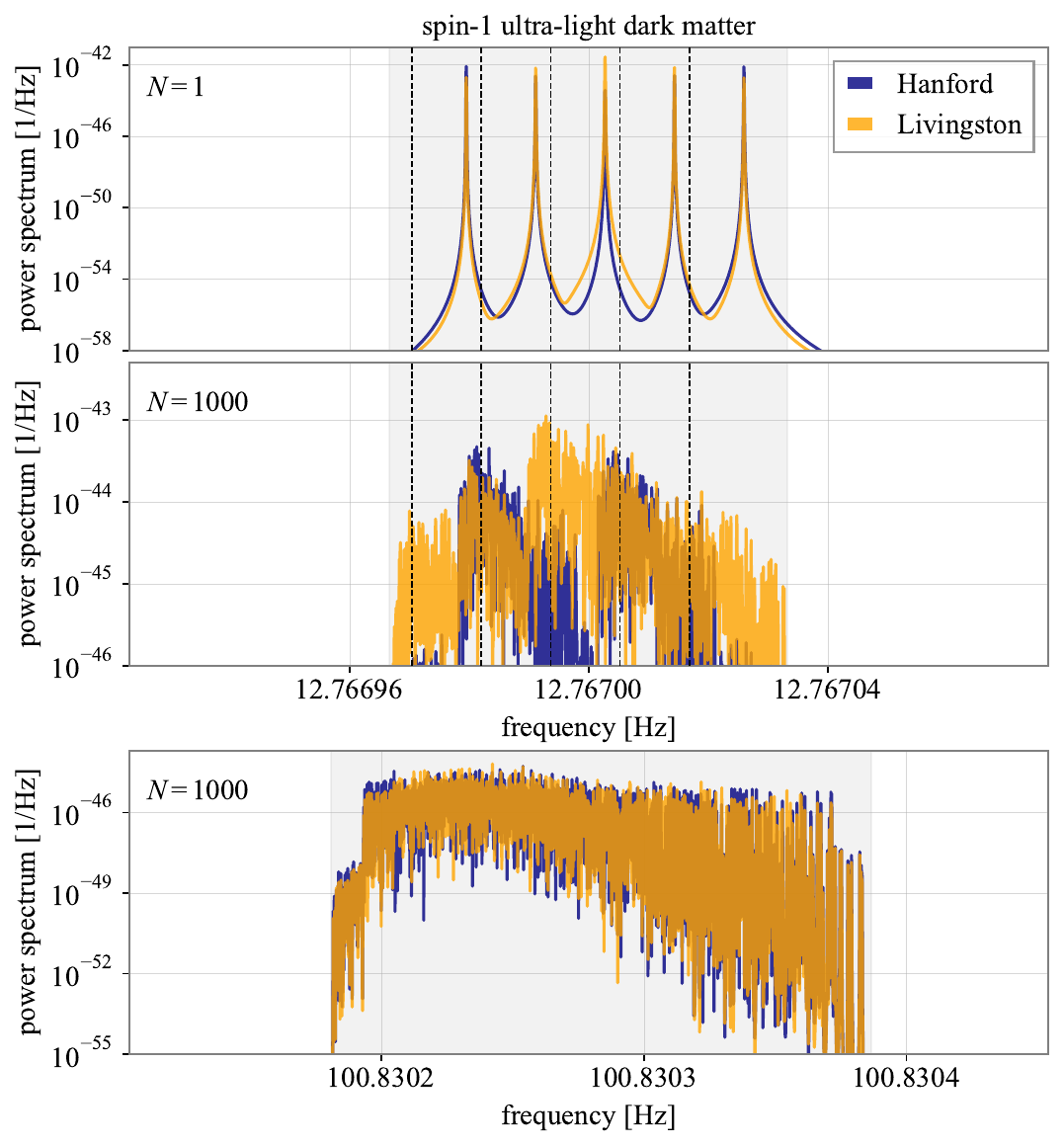}
\end{center}
\caption{Power spectra obtained from the simulation of $N$ spin-1 ultra-light dark matter waves with $m\simeq5.28\times 10^{-14}$\,eV (top, middle) and $m\simeq4.17\times 10^{-13}$\,eV (bottom). The dashed lines on the top and middle panels indicate the analytically derived positions of the peaks due to the sidereal modulation, considering the central value $v_0$ of the MB distribution -- the actual position for each wave depends on the specific draw of the velocity \(v_a\) from \cref{eq:MBPDF}. The gray region represents the theoretical prediction for the range that the peaks can occupy, i.e.\ from $f-\Omega_r/\pi$ to $f[1 + v_{\rm esc}^2/(2c)]+\Omega_r/\pi$. We do not report the spectrum of the high-frequency single wave since it has a similar spectral shape as the top panel. The frequency resolution of the spectra is equal to $4.89 \times 10^{-8}$~Hz.
}
\label{fig:ffts1}
\end{figure}

\begin{figure}
\begin{center}
\includegraphics[width=0.9\linewidth]{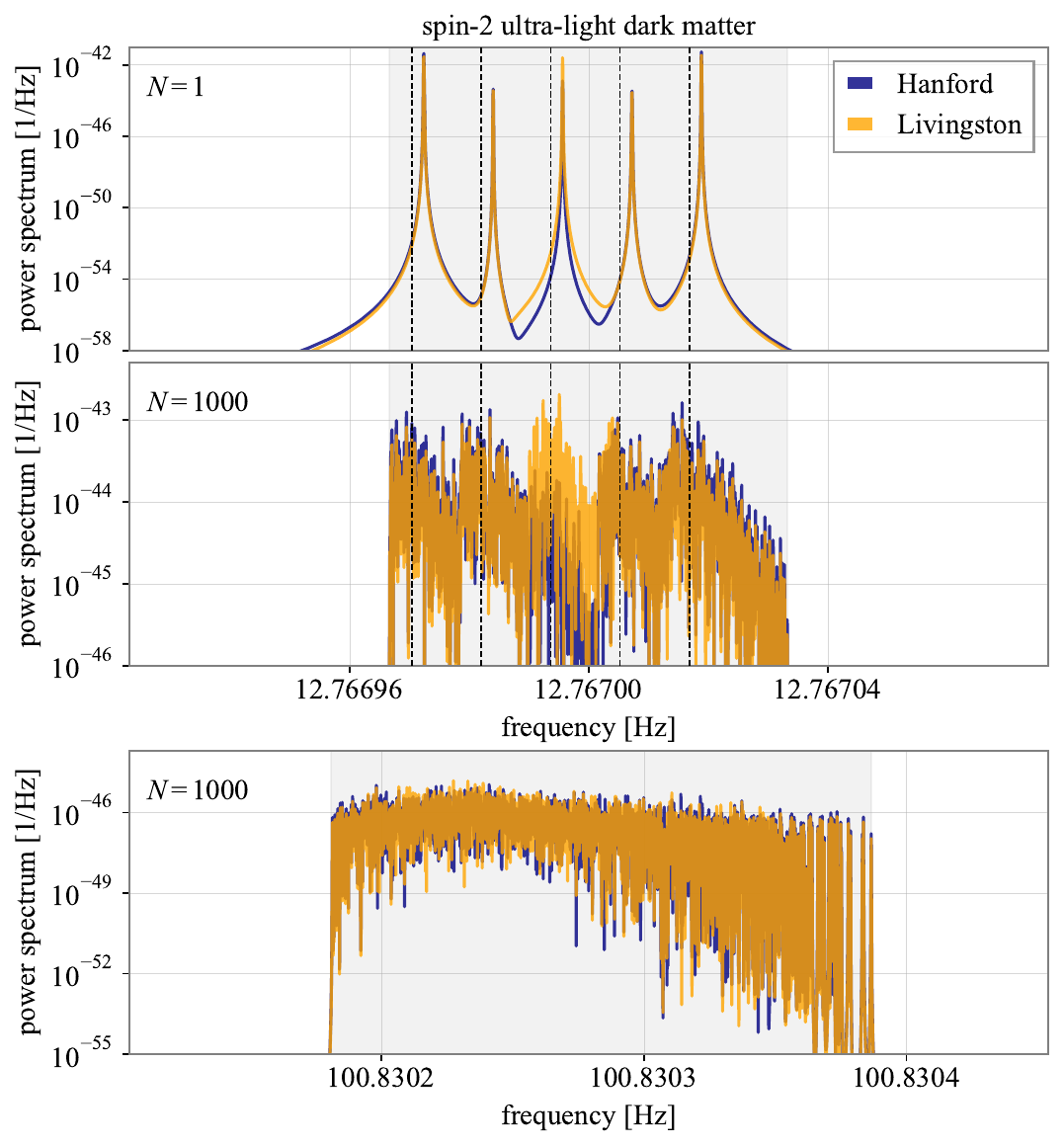}
\end{center}
\caption{Same as \cref{fig:ffts1} but for a simulated spin-2 signal.
}
\label{fig:ffts2}
\end{figure}

We show this noiseless strain in \cref{fig:strain} for the spin-1 and spin-2 cases. The corresponding power spectra in the frequency domain are shown in \cref{fig:ffts1} and \cref{fig:ffts2}. Both signals are simulated with equivalent couplings, meaning that a single spin-1 wave would lead to the same strain amplitude of the corresponding spin-2 wave, apart from the different responses and polarisations. The latter, together with the interferences that arise from superposing many waves, results in different strain amplitudes between the spin-1 and spin-2 scenarios as observed in \cref{fig:strain}, \cref{fig:ffts1} and \cref{fig:ffts2}~\cite{Jain:2021pnk}. 

In \cref{fig:ffts1} and \cref{fig:ffts2} we explicitly show the effects of the sidereal modulation and the MB frequency broadening in the spin-1 and spin-2 spectra, respectively. In the top and middle panels of both figures, we show an injected signal at \(f \approx 12.77\,\mathrm{Hz}\) $(m\simeq5.28\times 10^{-14}$\,eV), where we can observe the five peaks that arise due to the sidereal modulation. The spectrum of an individual wave ($N=1$) in the top panel clearly shows the peaks' structure (within the stochastic fluctuations due to the sampling of angles, polarisations and phase), while the superposition of $N=1000$ waves in the middle panel is what we would observe in reality. Comparing the middle panel of both figures, we see that the peaks located at $\omega=m\pm 2\Omega_r$ are subdominant in the spin-1 spectrum, while very prominent in the spin-2 spectrum, as expected from the fact that such peaks are related to $h_{\rm s}$ in the spin-1 strain. As we move to lower frequency, the hierarchy between the time and space spin-1 peaks becomes less pronounced, and it is in fact reversed below \(f \approx 9\,\mathrm{Hz}\). We also see that the peaks' amplitudes depend on the detector orientation, leading to different spectrum morphologies in Hanford and Livingston. 
In the lower panel instead we simulate a signal at \(f\approx100.83\,\mathrm{Hz}\) $(m\simeq4.17\times 10^{-13}$\,eV), for which MB frequency broadening, as expected, completely swallows the multiple peaks.

The signals considered here, i.e., ultra-light dark matter directly interacting with the detector, are just one type of quasi-monochromatic signal detectable by Earth-based detectors. Interferometers could also detect continuous gravitational waves from various astrophysical sources, including other dark matter candidates like superradiant boson clouds formed around spinning black holes (see e.g.~\cite{Piccinni_searches_2022,Miller:2025yyx} for a review). For the standard scenario of an isolated, spinning triaxial neutron star, sidereal modulation splits the signal into five frequencies~\cite{Jaranowski:1998qm}: $f$, $f \pm \Omega_r/(2\pi)$, $f \pm 2\Omega_r/(2\pi)$, while additional Doppler modulation contributes to additional spread. Here $f$ is the instantaneous gravitational frequency at $t=0$ at the solar system barycenter. The presence of a significant Doppler modulation may provide a distinguishing signature between ultra-light dark matter signals and continuous gravitational waves emitted by isolated spinning neutron stars. However, for sources near the ecliptic poles, the modulation is reduced and can be comparable to the intrinsic linewidth of ultra-light dark matter signals.
A similar consideration can be done for the case of continuous-wave signals coming from binaries, where a more complex forest of peaks is present due to additional orbital Doppler modulation at the binary orbital frequency. 
Of course all these considerations need to take into account the different sources of noise, but we expect that, at least in the case of a strong detection (high SNR), the amount of peaks, their relative amplitudes, as well as the separation between them, can still be visible and help discriminate between different source candidates. Finally, it is worth mentioning that other possibilities to distinguish the spin of ultra-light dark matter have been explored. In~\cite{Manita:2023mnc} the authors propose using the overlap reduction functions (ORF) of multiple detectors, while in~\cite{Miller:2022wxu} the authors propose the use of a Wiener filter. 

%%%%%%%%%%%%%%%%%%%%%%%%%%%%%%%%%%%%%%%%%%%%
\section{Novel search strategy}
\label{sec:method}
%%%%%%%%%%%%%%%%%%%%%%%%%%%%%%%%%%%%%%%%%%%%

Several searches have targeted ultra-light dark matter signals in the LVK detectors during the first four observing runs (O1–O4)~\cite{Guo:2019ker,LIGOScientific:2021ffg,PhysRevLett.133.101001,PhysRevD.110.042001,LIGOScientific:2025ttj} or using GEO600~\cite{Vermeulen:2021epa}.
Given the stochastic nature of these dark matter signals, cross-correlation methods are optimal for their detection, as they effectively suppress uncorrelated features~\cite{Pierce:2018xmy}. On the other hand, alternative approaches have been developed to overcome some of the technical limitations of the \sftxcorr method applied in~\cite{LIGOScientific:2021ffg}, which is sub-optimal for specific frequency (or mass) ranges due to the use of fixed integration times.
These methods, namely \bsdexcess~\cite{Miller:2020vsl} and \lpsd~\cite{gottel2025fastprecisespectralanalysis,Gottel_2024,TROBS2006120}, are based on excess-power analysis. 
Despite being less robust against noise, if compared to the \sftxcorr method,
these approaches show improved sensitivity, thanks to the use of optimal integration time in the setup of their searches.

In this section, we propose a new alternative approach that leverages the flexibility of the BSD framework to compute a cross-correlation statistics. In practice, we propose an improved cross-correlation implementation that ensures the optimal integration time is used across the analysed frequency bands (or dark matter masses).

In the following, we discuss the optimal search integration time, describe the implementation of the cross-correlation statistics within the \bsdxcorr approach and estimate its sensitivity. 

%%%%%%%%%%%%%%%%%%%%%%%%%%%%%%%%%%%%%%%%%%%%
\subsection{Optimal search integration time}\label{sec:tfftoptimal}

As discussed in \cref{sec:siderealmodulation} and \cref{sec:distinguish}, the signal will be impacted by the sidereal modulation, causing a frequency split of the signal into multiple frequency peaks. In addition to this effect, the signal is further modulated due to the MB distribution of velocities of the dark matter particles.  

\begin{figure}
\begin{center}
\includegraphics[width=0.7\linewidth]{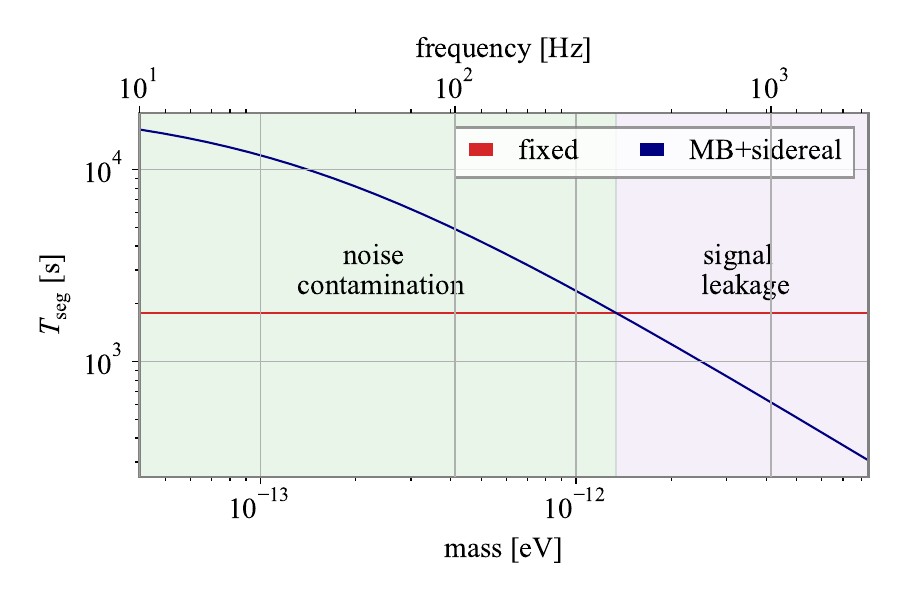}
\end{center}
\caption{Integration time $T_{\rm seg}$ as a function of the mass of the ultra-light particle. In the \bsdxcorr method, we adapt the integration time to its optimal value for each frequency (blue curve, see \cref{eq:tfftmax}). The \sftxcorr approach in~\cite{LIGOScientific:2025ttj}, uses a fixed value of $T_{\rm seg}=1800$~s for the entire frequency range (red curve). Smaller-than-optimal values will reduce the SNR due to noise contamination. On the other hand, when the integration time is larger-than-optimal, part of the signal is cut out of the bin. The result is a sub-optimal sensitivity for all the bins, except for the one where $T_{\rm seg}=T_{\rm seg}^{\rm opt}$.} \label{fig:Tfftccvsccbsd}
\end{figure}

\begin{figure}
\begin{center}
\includegraphics[width=0.7\linewidth]{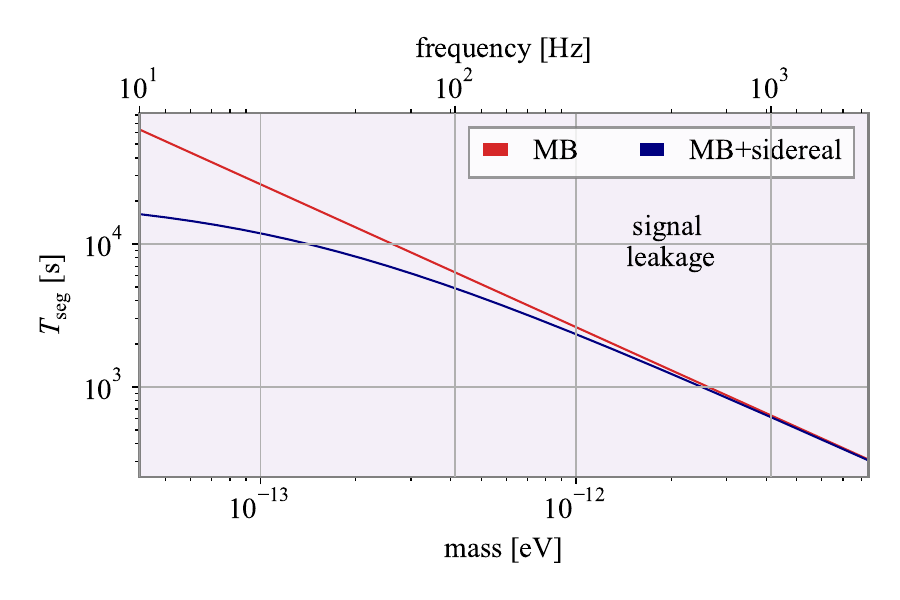}
\end{center}
\caption{Integration time $T_{\rm seg}$ as a function of the mass of the ultra-light particle. The red curve indicates the integration time $T_{\rm seg}^{\rm MB}(f)$ used in excess-power methods, while the blue curve corresponds to the optimal integration time $T_{\rm seg}^{\rm opt}(f)$. The sidereal modulation causes an extra broadening of the signal frequency, which, if not taken into account, results in a reduced bin width and the consequent signal leakage.} \label{fig:TFFTpeakeffect}
\end{figure}

Given the narrowband nature of the signal, the optimal strategy is to maximise the SNR by matching the search integration time ($T_{\rm seg}$) to the signal coherence time such that the broadened signal is confined within a single frequency bin $\Delta f\doteq1/T_{\rm seg}$. If the search integration time is larger than the signal coherence time, the signal power may be split in nearby frequency bins, causing signal leakage. On the other hand, if the integration time is too short, it yields wide bins that incorporate substantial noise, degrading sensitivity. Therefore, there is an optimal value of the search integration time, which we denote as $T_{{\rm seg}}^{\rm opt}(f)$, for each frequency given by
\begin{align}\label{eq:tfftmax}
    T_{\rm seg}^{\rm opt}(f) = \frac{1}{\Delta f^{\rm opt}(f)}, \quad\mathrm{with~} \Delta f^{\rm opt}(f) \doteq \Delta f_v(f)+\Delta f_{\rm sid}.
\end{align}
In \cref{fig:Tfftccvsccbsd}, we show the impact of the integration time on the search SNR by comparing a fixed choice, $T_{\rm seg}=1800\,\mathrm{s}$, adopted in the \sftxcorr\ method (fixed), with the frequency-dependent optimal integration time, $T_{\rm seg}^{\rm opt}(f)$, used in the \bsdxcorr\ approach (MB+sidereal). 
In \cref{fig:TFFTpeakeffect}, we further compare the integration time $T_{\rm seg}^{\rm MB}(f)$ employed in excess-power methods (MB) with the optimal choice $T_{\rm seg}^{\rm opt}(f)$. Details on the different search methods are provided in the following sections.

For reference, we provide the values of the optimal search integration time $T_{\rm seg}^{\rm opt}(f)$ and frequency bin $\Delta f^{\rm opt}(f)$ corresponding to three different masses $m$, or frequency $f$, in \cref{tab:cohtime}. 

\begin{table}[ht]
\centering
\begin{tabular}{cccc}   
%\hline
\toprule
$m$ [eV] & $f$ [Hz] & $T_{\rm seg}^{\rm opt}$ [s] & $\Delta f^{\rm opt}$ [Hz] \\ \midrule
$10^{-13}$\,  & $24.18$ & $11816.85$\, & $8.46\times 10^{-5}$\,  \\ %\hline
$10^{-12}$\,  & $241.80$ & $2333.38$\,  & $4.28 \times 10^{-4}$\, \\ %\hline
$10^{-11}$\,  & $2417.99$ & $258.54$\,   & $3.87\times 10^{-3}$\,  \\
%\hline
\bottomrule
\end{tabular}
\caption{Search integration time and frequency bin corresponding to specific ultra-light dark matter masses, and their corresponding frequencies.
}
\label{tab:cohtime}
\end{table}

Finally, we note that in the LVK searches employing the \bsdexcess method~\cite{LIGOScientific:2021ffg,LIGOScientific:2025ttj}, an integration time of $1.50~T_{\rm seg}^{\rm MB}(f)\doteq 1.50/\Delta f_v(f)$ has been used because the escape velocity $v_{\rm esc}$ is unlikely to be drawn from the MB distribution (physically, this corresponds to the fact that most of the waves that physically drive the signal will have much smaller velocities than the escape velocity), leading to a narrower MB spread compared to \cref{eq:MBestimate}. 

%%%%%%%%%%%%%%%%%%%%%%%%%%%%%%%%%%%%%%%%%%%%
\subsection{\bsdxcorr method}

Different gravitational wave detectors should experience nearly the same ultra-light dark matter signal, due to the fact that the coherence length of the field is much larger than the detectors' separation. For this reason, we can employ correlations between different detectors to search for ultra-light dark matter. 

In the standard \sftxcorr method used in~\cite{LIGOScientific:2021ffg}, the value of the search integration time is fixed to $T_{\rm seg}=1800~{\rm s}$, which corresponds to the default values of the LIGO SFT data~\cite{LSC_sft_2022}. As observed in \cref{sec:tfftoptimal}, this value is not optimal for the full frequency range analysed.
Such a limiting choice impacts the efficiency of the cross-correlation statistics.

To fully exploit the potential of correlation techniques, we propose a new implementation that combines the robustness of cross-correlation statistics with the flexibility of the BSD framework.
Direct access to the detector time series in the BSD format allows the integration time to be selected as a function of frequency. When matched to the coherence time of the ultra-light field, this choice enhances the signal power while suppressing uncorrelated noise, thereby ensuring optimal sensitivity across the full frequency range. 

In the following we detail the implementation of the cross-correlation statistics within our proposed \bsdxcorr method and estimate its corresponding theoretical sensitivity.

\subsubsection{Implementation of the cross-correlation statistics} 
\label{sec:BSDcross_correlation}

We describe the frequency-domain cross-correlation estimator between pairs of detectors using the strain time series in the BSD framework.
The data streams from the two detectors are first restricted to their common observation time and divided into consecutive segments. The analysis is performed in frequency sub-bands labelled by an index $b$, within which the Fast-Fourier Transform (FFT) duration $T_{\rm seg}^{(b)}$ is kept constant. The choice of $T_{\rm seg}^{(b)}$ is frequency dependent and is motivated by the expected coherence time of ultra-light dark matter signals, which varies across the analysed frequency range.

For each segment $i$, we compute the discrete Fourier transform coefficients of the strain data from detector $k=1,2$, denoted by $\tilde{z}_k^{(i)}(f)$. The corresponding one-sided power spectral densities (PSDs), $S_{n,k}^{(i)}(f)$, are estimated using a median-based procedure within coarse-grained frequency intervals.

The cross-correlation estimator is constructed, for each frequency bin $f_j \in b$, as 
\begin{equation}
\label{eq:sj}
C(f_j) = \frac{4}{N_{\rm seg}^{(b)} T_{\rm seg}^{(b)2}}
\sum_{i=1}^{N_{\rm seg}^{(b)}}
\frac{\tilde{z}_1^{(i)}(f_j)\,\tilde{z}_2^{(i)*}(f_j)}
{S_{n,1}^{(i)}(f_j)\,S_{n,2}^{(i)}(f_j)},
\end{equation}
where $^*$ denotes complex conjugation, $N_{\rm seg}^{(b)} \simeq \mathcal{O} T_{\rm obs} / T_{\rm seg}^{(b)}$ 
is the number of segments corresponding to band $b$ for the total observing time of $T_{\rm obs}$, and $\mathcal{O}=1$ and $\mathcal{O}=2$ indicate absence of overlap and $50\%$ overlap, respectively.
Assuming stationary, uncorrelated Gaussian noise between detectors, the variance of the estimator is given by
\begin{equation}
\label{eq:variance}
\sigma^2(f_j) = \frac{2}{N_{\rm seg}^{(b)\,2} T_{\rm seg}^{(b)2}} 
\sum_{i=1}^{N_{\rm seg}^{(b)}}
\frac{1}{S_{n,1}^{(i)}(f_j)\,S_{n,2}^{(i)}(f_j)},
\quad f_j \in b.
\end{equation}
The SNR is then computed as 
\begin{equation}
\label{eq:SNRcc}
\mathrm{SNR}(f_j) = \frac{C(f_j)}{\sigma(f_j)},
\quad f_j \in b.
\end{equation}
In the absence of a correlated signal, the SNR is expected to have zero mean, while the presence of an ultra-light dark matter signal produces a non-zero offset due to its coherent contribution across detectors.

The estimator above assumes unit overlap
reduction function (ORF); for a network of spatially separated detectors the ORF
$\gamma(f)$ encoding the geometric suppression can be straightforwardly
incorporated as a multiplicative weight on each cross-spectrum in \cref{eq:sj}.

\begin{figure}
\begin{center}
\includegraphics[width=1.0\linewidth]{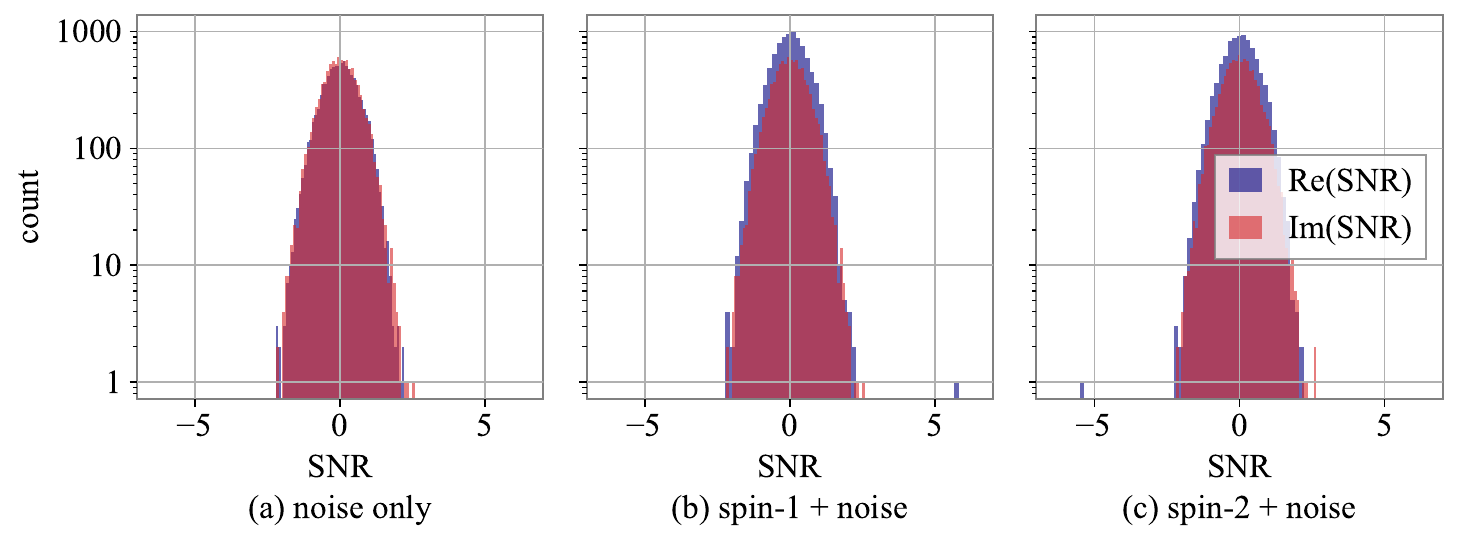}
\end{center}
\caption{SNR histogram of the \bsdxcorr statistics computed between the two LIGO's Livingston and Hanford detectors in the $107~{\rm Hz}-108~{\rm Hz}$ band. 
The left panel (a) shows the SNR distribution obtained in noise only, while the middle panel (b) shows the distribution obtained after injecting a spin-1 signal in real noise, with coupling $\epsilon_D=3.79\times 10^{-23}$, $m=4.44\times 10^{-13}$eV; the right panel (c) shows the case for the spin-2 signal, with $\alpha=6.00\times10^{-8}$ and same mass as the spin-1 case. The statistic has been computed using the optimal $T_{\rm seg}^{\rm opt}$ for that frequency band and no overlap between segments. The outliers present in the middle and right panels, with $|{\rm Re}({\rm SNR})|>5$,  correspond to the injected signals in the two cases.
}
\label{fig:hist}
\end{figure}

In \cref{fig:hist} we report the histograms of the SNR of the cross-correlation statistics obtained by the \bsdxcorr approach. We used data from the Livingston and Hanford detectors for the cases of (a) noise only, (b)  spin-1 injection, and (c) spin-2 injection. Both injections are performed in real data with a particle mass of $4.44\times 10^{-13}$~eV,  corresponding to a frequency of $107.36$~Hz in the frequency band $b$ of $[107-108]$ Hz. As for the couplings, we have $\epsilon_D=3.79\times 10^{-23}$ for spin 1 and $\alpha=6.00\times10^{-8}$ for spin 2. 
The data span GPS times 1368979895~s to 1389428369~s, from  the O4a run, and the signal is injected over the full observation time $T_{\rm obs}$. The corresponding $T_{\rm seg}^{(b)}$ for the considered injection frequency is $T_{\rm seg}=4627.79\,\mathrm{s}$, chosen as the optimal integration time $T_{\rm seg}^{\rm opt}$ given by \cref{eq:tfftmax}. Without loss of generality, in the following we omit the superscript $b$ and assume that the integration time varies across frequency bins. We note, however, that in a realistic search the frequency range may need to be partitioned into broader sub-bands, indexed by $b$.

\subsubsection{Sensitivity of the \bsdxcorr method}

\begin{figure}
\begin{center}
\includegraphics[width=1.0\linewidth]{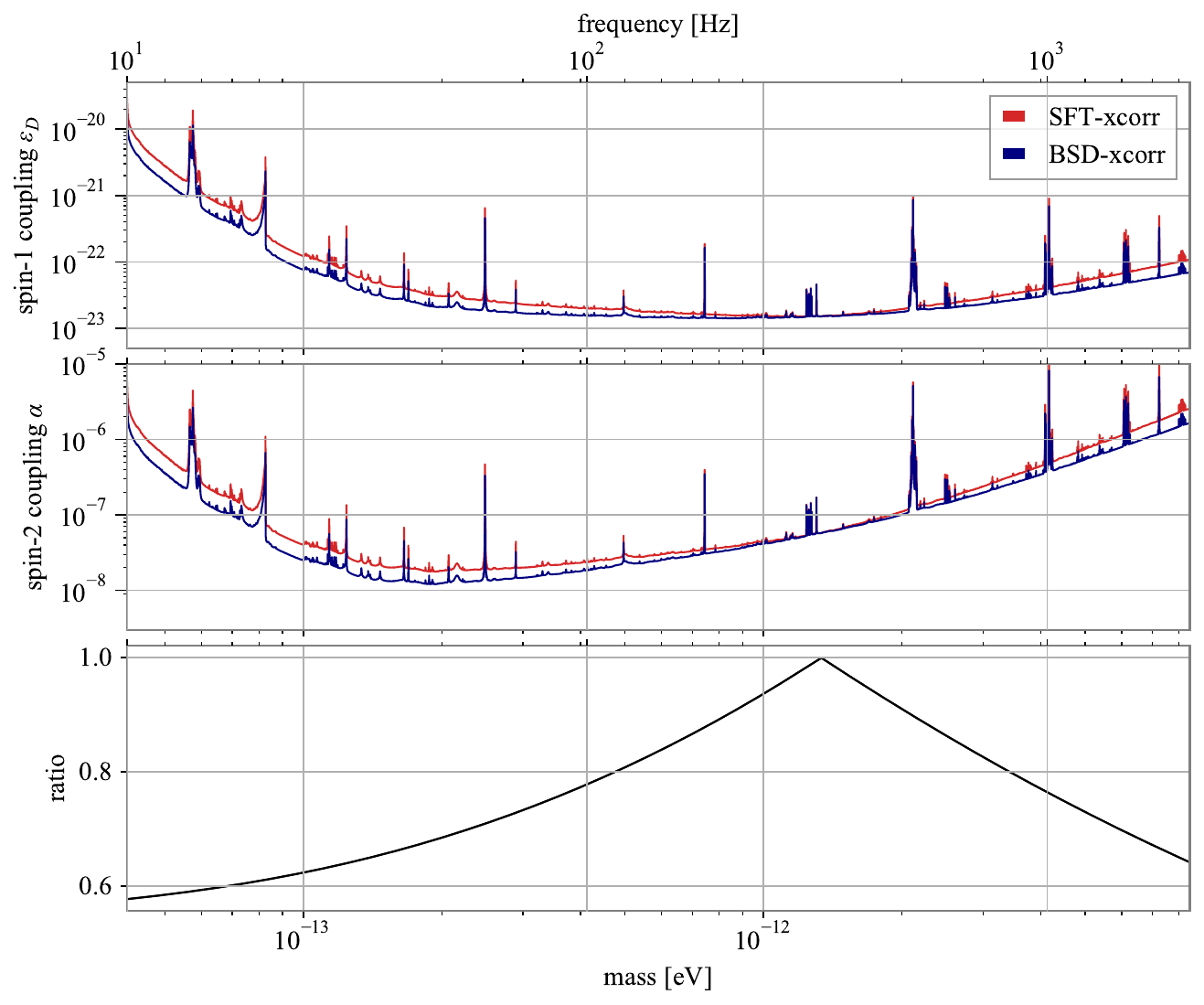}
\end{center}
\caption{Theoretical sensitivities (non-overlapping segments), of the \sftxcorr and \bsdxcorr methods by requiring ${\rm SNR}=5.00$. The red curve indicates the sensitivity of \sftxcorr using $T_{\rm seg}=1800~$s for all bins, while the blue curve corresponds to the sensitivity of \bsdxcorr using the optimal integration time $T_{\rm seg}^{\rm opt}(f)$ across the frequency range. The top (middle) panel shows the spin-1 (spin-2) results, while the bottom panel shows the ratio between the \bsdxcorr and \sftxcorr  curves, which is equivalent for both spins.
\label{fig:thsensCCvsCCBSD} }
\end{figure}

Let us now estimate the theoretical sensitivity of the \bsdxcorr method. To do so we need to model the discrete Fourier transform of the data strain $z(t)= n(t)+ h(t)$ here written as the sum of noise and signal.  
Following the treatment of~\cite{PhysRevD.90.042002}, the Fourier transform of a narrowband finite length signal can be written as 
\begin{equation}
\label{eq:hfourier}
    \tilde{h}(f_j) 
    =\frac{T_{{\rm seg}}(f_j)}{2}\sqrt{2\left<h^2\right>}\sqrt{\frac{2.43}{\pi}},
\end{equation}
where $2\left<h^2\right>=\left<h^2\right>_{v,k,\varepsilon}$ is applied to remove the time average, which is accounted for in the frequency average factor $\sqrt{2.43/\pi}$ in frequency domain. Within the assumption of uncorrelated noise, we approximate $\tilde{z}_1^{(i)}(f_j)=\tilde{z}_2^{(i)}(f_j)\simeq \tilde{h}(f_j)$, such that the SNR from \cref{eq:SNRcc} reads
\begin{align}\label{eq:SNRbasic}
    {\rm SNR}(f_j) =|\tilde{h}(f_{j})|^2 \frac{2\sqrt{2N_{\rm seg}(f_j)}}{T_{\rm seg}(f_j)}\left[\Bigl<\frac{1}{ S_{n,1}^{(i)}(f_j) S_{n,2}^{(i)}(f_j) }\Bigl>\right]^{\frac{1}{2}},
\end{align}
where $<\cdot>$ denotes average over $N_{\rm seg}$ time segments,  while $\tilde{h}(f_j)$ is related to $\sqrt{\left<h^ 2\right>}$ via \cref{eq:hfourier}.
In the \bsdxcorr implementation, we require $T_{{\rm seg}}(f_j)=T_{{\rm seg}}^{\rm opt}(f_j)$ for all the $j$ bins, ensuring a nearly optimal SNR recovery across all frequencies.
The corresponding minimum detectable strain is then computed as
\begin{align}\label{eq:sensbsdxcorr}
    \sqrt{\left<h^2\right>}_{\rm min} = \left[\frac{{\rm SNR}(f_j)}{ T_{{\rm seg}}(f_j)}\frac{\pi}{2.43}\right]^{\frac{1}{2}} \left[\frac{1}{2N_{\rm seg}(f_j)}\right]^{\frac{1}{4}}\left[\Bigl<\frac{1}{S_{n,1}^{(i)}(f_j)S_{n,2}^{(i)}(f_j)}\Bigl>\right]^{-\frac{1}{4}},
\end{align}
while the minimum detectable couplings $\epsilon_D$ and $\alpha$ (corresponding to a SNR=5.00) are obtained via \cref{eq:avrstrainspin1} and \cref{eq:avrstrainspin2} respectively, and are shown as the blue curves in the top and middle panels of \cref{fig:thsensCCvsCCBSD}, with $M/Q_D=m_p$ chosen as the proton mass. 
We note that, when implementing the \bsdxcorr method with $50\%$ overlap between chunks, the effective number of Fourier transforms doubles, resulting in a sensitivity improvement of $16\%$ (i.e.\ a reduction of the minimum detectable strain by a factor of $0.84$). 

\section{Impact on existing search strategies}\label{sec:impact}
In the following we quantify the impact of the use of sub-optimal integration times when searching for dark matter with the \sftxcorr~\cite{Pierce:2018xmy}, \bsdexcess~\cite{Miller:2020vsl}, and \lpsd~\cite{gottel2025fastprecisespectralanalysis, TROBS2006120} methods applied in recent searches.  
We estimate the sensitivity improvements by modelling the effects of signal leakage and noise contamination in the corresponding theoretical sensitivities, 
and report projected sensitivities obtained by: (i) using the cross-correlation statistics within the \bsdxcorr approach, opposed to the \sftxcorr method; (ii) accounting for sidereal modulation in the \bsdexcess and \lpsd methods.

\subsection{\sftxcorr}

In this section we evaluate the improvement reached by the \bsdxcorr method compared to the previously implemented \sftxcorr.  The sensitivity computed in \cref{eq:sensbsdxcorr} assumes that the signal power is fully captured in the $j$-th frequency bin, which does not hold when the chosen search integration time  $T_{\rm seg}(f_j)$ is not optimal. 
We can easily generalise the sensitivity to model the captured signal power as a function of the integration time by introducing the factor $\eta(f_j)$
\begin{equation}
\label{eq:eta}
\eta(f_j) \doteq
\begin{cases}
1, & T_{{\rm seg}}^{\rm opt} (f_j)\geqslant T_{{\rm seg}} \\
\frac{T_{{\rm seg}}^{\rm opt}(f_j)}{T_{{\rm seg}}}, & T_{{\rm seg}}^{\rm opt}(f_j) < T_{{\rm seg}}
\end{cases},
\end{equation} 
such that the theoretical sensitivity of the \sftxcorr method is obtained as
\begin{align}
    \sqrt{\left<h^2\right>}_{\rm min} = \left[\frac{{\rm SNR}(f_j)}{\eta(f_j) T_{{\rm seg}}}\frac{\pi}{2.43}\right]^{\frac{1}{2}} \left[\frac{1}{2N_{\rm seg}}\right]^{\frac{1}{4}}\left[\Bigl<\frac{1}{S_{n,1}^{(i)}(f_j)S_{n,2}^{(i)}(f_j)}\Bigl>\right]^{-\frac{1}{4}},
\end{align}
where $T_{\rm seg}$ and $N_{\rm seg}$ without a frequency dependence denote the fixed values employed in this approach.
Therefore, when $T_{\rm seg} > T_{\rm seg}^{\rm opt}(f_j)$, the effective signal recovery is suppressed by the factor $\eta(f_j) \approx T_{\rm seg}^{\rm opt}(f_j)/T_{\rm seg}$, so that the SNR scales with $T_{\rm seg}^{\rm opt}(f_j)$. In this regime, extending the integration time beyond the optimal value does not increase the coherent signal contribution, but instead spreads it over a longer duration, effectively reducing its recoverable power.
In contrast, when $T_{\rm seg} < T_{\rm seg}^{\rm opt}(f_j)$, the SNR scales with $T_{\rm seg}$. Although no loss is incurred from $\eta(f_j)$, the integration time is insufficient to fully accumulate the signal power, and the corresponding frequency bin remains contaminated by noise. Consequently, using a fixed $T_{\rm seg}$ across all frequency bins $j$ leads to a loss of sensitivity in general, with optimal sensitivity  achieved only for the bin where $T_{\rm seg} = T_{\rm seg}^{\rm opt}(f_j)$, as shown in \cref{fig:Tfftccvsccbsd}.

We can express the sensitivity in terms of the minimum detectable couplings $\epsilon_D$ and $\alpha$ by means of \cref{eq:avrstrainspin1} and \cref{eq:avrstrainspin2}. 
Such curves are presented in \cref{fig:thsensCCvsCCBSD} for the \bsdxcorr and the standard \sftxcorr methods, where the latter employs a fixed search integration time of $T_{\rm seg}=1800$~s for all frequencies.  
The improvement observed at low frequencies is due to the reduction of noise contamination in the frequency bin, while at high frequencies signal leakage is prevented.
The improvement ratio shown in the bottom panel is computed as
\begin{align}\label{eq:rbsdxcorr}
r_{\rm BSD\text{-}xcorr} = \eta(f_j)^{\frac{1}{2}}
\left(\frac{T_{\rm seg}}{T_{{\rm seg}}^{\rm opt}(f_j)}\right)^{\frac{1}{4}}.
\end{align}

We note that in~\cite{Morisaki:2025bjs} the authors introduce another detection statistic, named coherent SNR, showing that it is possible to optimally extract ultra-light bosonic dark matter signals from data segments of arbitrary duration by exploiting their expected correlations. It is then shown that Fourier-domain data from segments of arbitrary length can be coherently combined through a weighted sum of inter-segment correlations, with the weights encoding the expected signal coherence based on a 
MB velocity distribution, but without accounting for the separation of peaks due to the sidereal modulation obtained in the present work.

\subsection{Excess-power methods}
Lastly, we estimate the sensitivity improvements of the excess-power methods. Without loss of generality, we follow the derivation of~\cite{PhysRevD.90.042002}, with corrections from~\cite{Palomba2025_peakmap_sensitivity} (see \cref{app:bsd}), to compute the \bsdexcess sensitivity and rescale  it to the \lpsd case.

The minimum detectable strain for a semi-coherent excess-power search scales as
\begin{align}\label{eq:bsdsensitivity}
\sqrt{\langle h^2 \rangle}_{\rm min}
\approx
\frac{\mathcal{B}}{N_{\rm seg}^{1/4}(f_j)}
\sqrt{\frac{S_n(f_j)}{\zeta(f_j)\,  T_{\rm seg}(f_j)}}  \propto  
\zeta(f_j)^{-\frac{1}{2}} T_{\rm seg}(f_j)^{-\frac{1}{4}}
\end{align}
where $\mathcal{B}$ is a numerical prefactor depending on the search implementation. We introduce
\begin{align}\label{eq:etapeaksep}
\zeta(f_j) \doteq \frac{T_{\rm seg}^{\rm opt}(f_j)}{\tau T_{\rm seg}^{\rm MB}(f_j)},
\end{align}
to quantify the  fractional sensitivity loss induced by neglecting the sidereal spectral structure, when the sub-optimal FFT length is chosen, as shown in \cref{fig:TFFTpeakeffect}. In the standard implementations of the excess-power methods, $T_{\rm seg}(f_j)=\tau T_{\rm seg}^{\rm MB}(f_j)$, 
while $\tau=1.50$ in the \bsdexcess case~\cite{LIGOScientific:2025ttj} and $\tau=1.00$ for \lpsd.

\begin{figure}
\begin{center}
\includegraphics[width=1.0\linewidth]{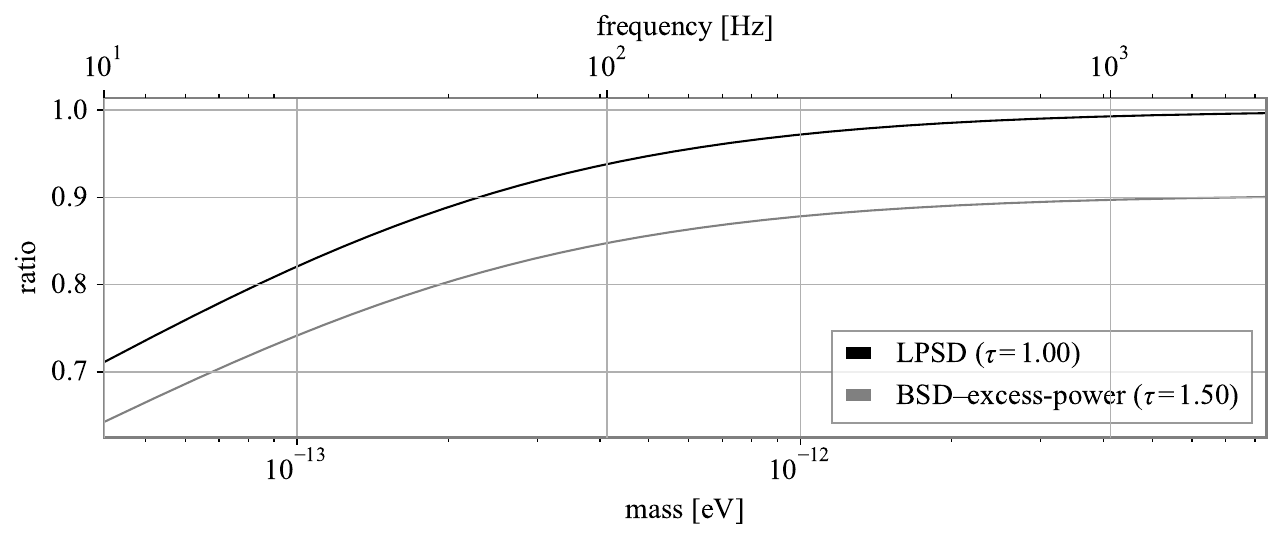}
\end{center}
\caption{Ratio between the theoretical sensitivities of the optimised excess-power methods (obtained using the optimal integration time $T_{\rm seg}^{\rm opt}(f)$) and the standard excess-power methods (which use only the integration time $\tau T_{{\rm seg}}^{\rm MB}(f)$). The grey curve corresponds to the \bsdexcess method improvement, with $\tau=1.50$, while the black curve corresponds to the \lpsd method improvement, with $\tau=1.00$.
\label{fig:BSDpeaksepornot} }
\end{figure}

The corresponding sensitivity gain between the optimised and standard excess-power methods is
\begin{align}\label{eq:rbsdlpsd}
r_{\rm BSD,LPSD}
=
\left(\frac{T_{\rm seg}^{\rm opt}(f_j)}{\tau\,T_{\rm seg}^{\rm MB}(f_j)}\right)^{1/4}.
\end{align}
As shown in \cref{fig:BSDpeaksepornot}, the improvement is most significant at low frequencies, where the sidereal broadening is comparable to or larger than the MB contribution.

%%%%%%%%%%%%%%%%%%%%%%%%%%%%%%%%%%%%%%%%%%%%
\subsection{Projections}
\label{ssec:project}

\begin{figure}
\begin{center}
\includegraphics[width=1.0\linewidth]{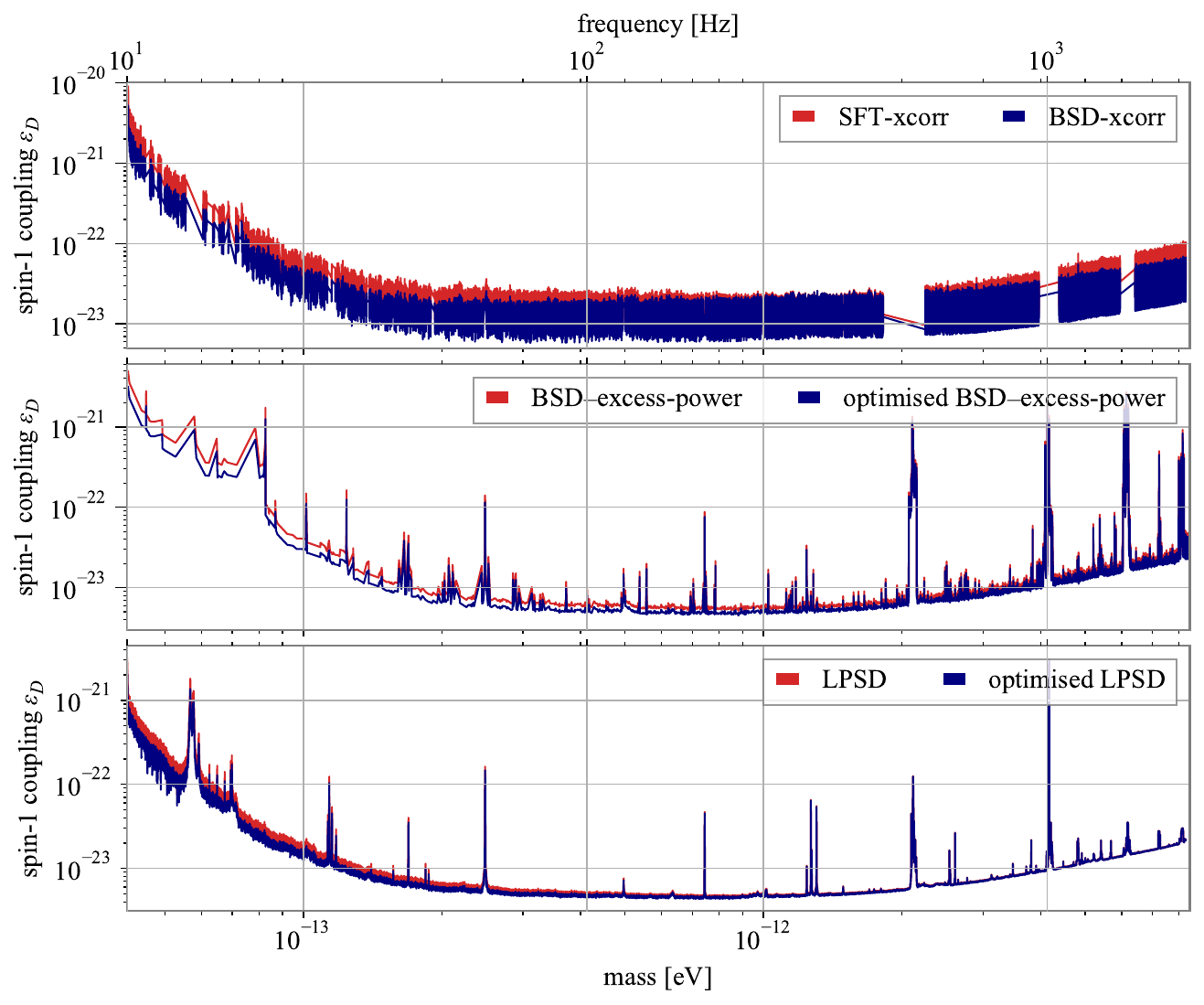}
\end{center}
\caption{Forecast of the LVK O4a search upper limits~\cite{LIGOScientific:2025ttj} for spin-1 ultra-light dark matter after implementing the optimal search integration time $T_{\rm seg}^{\rm opt}(f)$ proposed in the present work. The top panel shows the expected upper limit improvement of \bsdxcorr relative to \sftxcorr. The middle and bottom panels show the improvements of \bsdexcess and \lpsd methods, respectively, when implemented with the optimal integration time.
\label{fig:forecast1}}
\end{figure}

\begin{figure}
\begin{center}
\includegraphics[width=1.0\linewidth]{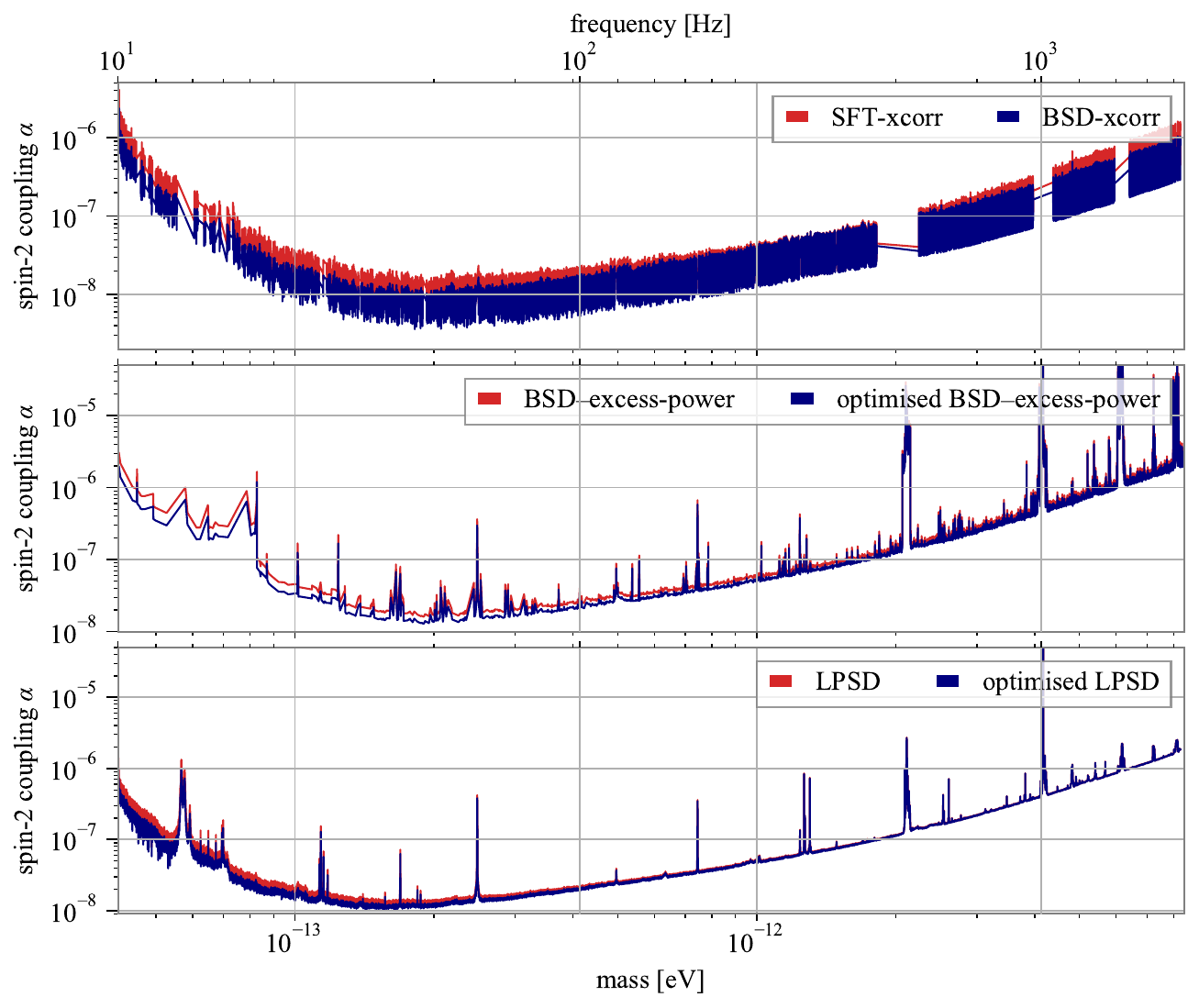}
\end{center}
\caption{Same as \cref{fig:forecast1} but for spin-2 ultra-light dark matter.
\label{fig:forecast2}}
\end{figure}

In \cref{fig:forecast1} (\cref{fig:forecast2}) we show the expected improvements on the existing LVK O4a bounds on the spin-1 (spin-2) coupling \(\epsilon_D\) of \cref{eq:Lintspin1} (\(\alpha\) of \cref{eq:Lintspin2}) for all three search methods. In the top panel we show the forecast for the new \bsdxcorr method (including the broadening caused by the sidereal modulation) compared to the existing limits from the \sftxcorr method. The middle and bottom panels show the expected sensitivity improvements for the \bsdexcess and \lpsd methods, respectively, when the sidereal modulation is taken into account. Notice that the \bsdexcess method adopted the longer search integration time of \(1.50 T^\mathrm{MB}_\mathrm{FFT}\), but not the \lpsd method; this leads to a more visible improvement in the \bsdxcorr case because now the ratio between \(\Delta f_\mathrm{sid}\) and \(\Delta f_v\) is 1.5~times larger across all frequencies. Note that for both spin-1 and spin-2 signals, the sensitivity improvements are identical, as they depend only on $\Delta f_{\mathrm{sid}}$, which is the same for both cases. 

To quantify the improvements on the upper limits of the coupling constants, we compute the ratio between the \bsdxcorr and \sftxcorr sensitivities given in \cref{eq:rbsdxcorr}, as well as the ratio (see \cref{eq:rbsdlpsd}) between the optimised excess-power methods (implemented using the optimal integration times reported by this work) and their standard implementations. These ratios are collected in \cref{tab:ratios}, where we also provide the relative improvement.

\begin{table}[ht]
\centering
\begin{tabular}{c c|ccc}
\toprule
$f\,[\mathrm{Hz}]$ & Quantity 
& \sftxcorr & \bsdexcess & \lpsd \\
\midrule

\multirow{2}{*}{10} 
 & Ratio            & $0.58$ & $0.64$ & $0.71$ \\
 & Improvement [\%] & $42$   & $36$   & $29$   \\
\hline

\multirow{2}{*}{100} 
 & Ratio            & $0.78$ & $0.85$ & $0.94$ \\
 & Improvement [\%] & $22$   & $15$   & $6.2$  \\
\hline

\multirow{2}{*}{2000} 
 & Ratio            & $0.65$ & $0.90$ & $1.0$  \\
 & Improvement [\%] & $35$   & $10$   & $0.36$ \\
\bottomrule
\end{tabular}
\caption{Optimised-to-existing bounds ratios for the three search methods, and their corresponding improvements in percentage, at three representative frequencies.}
\label{tab:ratios}
\end{table}

%%%%%%%%%%%%%%%%%%%%%%%%%%%%%%%%%%%%%%%%%%%%
\section{Conclusion and discussion}
\label{sec:end}
%%%%%%%%%%%%%%%%%%%%%%%%%%%%%%%%%%%%%%%%%%%%

Laser interferometers such as the LVK detectors, have emerged as among the most sensitive detectors of ultra-light dark matter in the audio frequency band, surpassing other dedicated experiments by up to several orders of magnitude. In this paper we build on this programme by proposing a new method, dubbed ``\bsdxcorr'', to search for narrowband, correlated and stochastic ultra-light dark matter signals in LVK data. This new method brings together the variable-integration-time implementation of the \bsdexcess and \lpsd methods, tailored to the optimal signal coherence time in order to contain the whole signal within one frequency bin, with the signal correlation properties across multiple detectors that is exploited in the cross-correlation search method. In practice, this is implemented by allowing for the integration time of the cross-correlation to vary across frequencies in order to match the signal coherence time.

In order to determine the optimal integration time we study the morphology of the signal and, for the first time, take into account the impact of the sidereal motion (and the detector orientation); we find that the spin-1 and spin-2 narrow-band peaks split into multiple ones, whose structure depends on the Lorentz nature of the detector response function, the spin of the signal and the position and orientation of the detector arms. In particular, we find that the signal produced by the spin-1 finite-time travelling effect generically exhibits three peaks at \(\omega=m,\,m\pm\Omega_r\), whereas the spin-1 differential strain (which is first-order in the field's gradient) as well as the spin-2 signal exhibit two farther peaks at \(\omega=m\pm2\Omega_r\). At low frequencies, because the finite-time travelling effect dominates, the spin-1 signal is primarily distributed among the three central peaks, whereas the spin-2 signal generically populates five. One more notable feature of these signals is that in all cases, owing to its specific arm orientation, the central peak observed in LIGO-Hanford (but not in LIGO-Livingston) will be suppressed with respect to the other peaks. At high frequencies the spin-1 field gradients dominate the response, so both spin-1 and spin-2 signals populate all five peaks equally (modulo the projection onto each detector's orientation); however, for frequencies higher than about \(30\,\mathrm{Hz}\) the multiple peaks merge within the MB frequency spread and become indistinguishable. These unique spectral features can help discriminating between signals from different dark matter models, as well as between dark matter and other continuous, nearly monochromatic sources such as neutron stars or emission from boson clouds.

Thanks to our optimised method, together with the new understanding of the signal morphology, the LVK sensitivity reach to ultra-light dark matter can be improved in two ways. Firstly, the sensitivity of the \sftxcorr method can be optimised via the \bsdxcorr method, which is expected to be more sensitive across all frequency bins, except for the one for which the fixed \sftxcorr integration time of \(1800\,\mathrm{s}\) corresponds to the optimal integration time. In particular, at high frequencies, the optimal integration time is shorter than the fixed integration time, and the signal will leak outside of a single bin; conversely, at low frequency, because the optimal integration time will be longer, noise will contaminate the signal we seek to detect (if we ignore the peak separation). Secondly, all variable-integration-time methods (\bsdxcorr, \bsdexcess and \lpsd) should take into account the peak separation as well as the MB spread when computing the optimal integration time for a search. Given that the effect of the sidereal modulation is always to generate multiple peaks, and therefore to widen the frequency range of the signal, ignoring it always leads to signal leakage, which is more pronounced at low frequency as that is where the MB width is comparable or smaller than the peak separation.

In order to quantify the impact of our optimised method, we forecast the sensitivities to both spin-1 and spin-2 ultra-light dark matter couplings by applying it onto the existing LVK O4a limits. In the case of the \bsdexcess (\lpsd) methods, the only improvement comes from taking into account the sidereal peak separation. We find that in this case the limits on the spin-1 and spin-2 couplings (\(\epsilon_D\) and \(\alpha\)), can improve by approximately 36\% (29\%) at \(10\,\mathrm{Hz}\) and 15\% (6\%) at \(100\,\mathrm{Hz}\) -- these correspond to optimised-to-existing ratios of 0.64 (0.71) and 0.84 (0.94) at \(10\,\mathrm{Hz}\) and \(100\,\mathrm{Hz}\), respectively. In the case of the \sftxcorr method, however, our optimised \bsdxcorr method (including the sidereal peak separation) can improve on those limits by approximately 42\% at \(10\,\mathrm{Hz}\), 22\% at \(100\,\mathrm{Hz}\) and back up at about 35\% at \(2\,\mathrm{kHz}\) -- corresponding to ratios 0.58, 0.78 and 0.65.

The LVK limits on the spin-1 ultra-light dark matter \(\epsilon_D\) coupling are the most competitive ones above approximately \(40\,\mathrm{Hz}\) up to \(2\,\mathrm{kHz}\), whereas at lower frequencies measurements from MICROSCOPE impose stricter limits. For spin-2 ultra-light dark matter instead, the LVK limits are the most competitive ones across the whole \(10\,\mathrm{Hz}\text{--}2\,\mathrm{kHz}\) frequency range, which is perhaps unsurprising given that laser interferometers were built to detect precisely the type of differential strain that a spin-2 dark matter coupled to standard matter produces. Thence, while our new method is expected to further improve the limits for both spin-1 and spin-2 dark matter, it is especially relevant for the spin-2 case at low frequency, where LVK could resolve the peak separation of the spin-2 signal and contrast this signal with other putative ones.

Looking into the future, the sidereal modulation effect that we have described in this work is particularly relevant because planned next-generation, ground-based detectors, such as the Cosmic Explorer and, in particular, the Einstein Telescope will extend the accessible low-frequency range to below \(10\,\mathrm{Hz}\) and down to \(1\,\mathrm{Hz}\). At such low frequencies we may be able to fully resolve the individual peaks, and each peak's MB spread alone would dictate the length of the optimal integration time. In this regime, thanks to the richness of the signal structure in frequency domain alone, we would be able to tell apart different types of continuous sources.

%%%%%%%%%%%%%%%%%%%%%%%%%%%%%%%%%%%%%%%%%%%%
\acknowledgments

We would like to thank Alexandre S. Göttel, Huaike Guo, Andrew L.\ Miller, Cristiano Palomba, Mai Qiao, and the LVK Continuous Wave group for valuable discussions that significantly contributed to the development of this work. This research has made use of data or software obtained from the Gravitational Wave Open Science Center (gwosc.org), a service of the LIGO Scientific Collaboration, the Virgo Collaboration, and KAGRA. This material is based upon work supported by NSF's LIGO Laboratory which is a major facility fully funded by the National Science Foundation, as well as the Science and Technology Facilities Council (STFC) of the United Kingdom, the Max-Planck-Society (MPS), and the State of Niedersachsen/Germany for support of the construction of Advanced LIGO and construction and operation of the GEO600 detector. Additional support for Advanced LIGO was provided by the Australian Research Council. Virgo is funded, through the European Gravitational Observatory (EGO), by the French Centre National de Recherche Scientifique (CNRS), the Italian Istituto Nazionale di Fisica Nucleare (INFN) and the Dutch Nikhef, with contributions by institutions from Belgium, Germany, Greece, Hungary, Ireland, Japan, Monaco, Poland, Portugal, Spain. KAGRA is supported by Ministry of Education, Culture, Sports, Science and Technology (MEXT), Japan Society for the Promotion of Science (JSPS) in Japan; National Research Foundation (NRF) and Ministry of Science and ICT (MSIT) in Korea; Academia Sinica (AS) and National Science and Technology Council (NSTC) in Taiwan. P.C.M.D.\ is supported by the Czech Science Foundation (GAČR) project PreCOG (Grant No.\ 24-10780S). O.J.P.\ is supported by the Spanish Ministerio de Ciencia, Innovacion y Universidades Ramon y Cajal, RYC2023-044489-I funded by MCIN/AEI/10.13039 /501100011033 and the FSE+ and cofinanced by the Universitat de les Illes Balears (UIB). This work was supported by UIB with funds from the Programa de Foment de la Recerca i la Innovació de la UIB 2024-2026 (supported by the yearly plan of the Tourist Stay Tax ITS2023-086); the Spanish Agencia Estatal de Investigación grants RED2024-153978-E, RED2024-153735-E, funded by MICIU/AEI/10.13039/501100011033 and the ERDF/EU; and the Comunitat Autònoma de les Illes Balears through the Conselleria d'Educació i Universitats with funds from the ERDF (SINCO2022/18146). F.U.\ acknowledges support from the European Structural and Investment Funds and the Czech Ministry of Education, Youth and Sports (project No. FORTE--CZ.02.01.01/00/22\_008/0004632). This article is based upon work from the COST Action COSMIC WISPers CA21106, supported by COST (European Cooperation in Science and Technology).

%%%%%%%%%%%%%%%%%%%%%%%%%%%%%%%%%%%%%%%%%%%%
%%%%%%%%%%%%%%%%%%%%%%%%%%%%%%%%%%%%%%%%%%%%
\appendix

%%%%%%%%%%%%%%%%%%%%%%%%%%%%%%%%%%%%%%%%%%%%
\section{Definitions}
\label{app:def}

The massive spin-1 field of \cref{sec:spin1uldm} is described, in the mostly positive metric convention which we adopt here, by the Proca lagrangian
\begin{align}
    {\cal L}_\mathrm{spin-1} = -\frac14 F_{\mu\nu} F^{\mu\nu} - \frac12 m^2 A_\mu A^\mu.
\end{align}
The massive spin-2 field of \cref{sec:spin2uldm} instead is described by the Fierz-Pauli lagrangian
\begin{align}
    {\cal L}_\mathrm{spin-2} = -M^{\mu\nu}{\mathcal E}_{\mu\nu}^{~~~\rho\sigma} M_{\rho\sigma}-\frac12 m^2\left(M_{\mu\nu}M^{\mu\nu}-M^2\right).
\end{align}
where $M \doteq \eta^{\mu\nu} M_{\mu\nu}$ is the trace of the field and ${\mathcal E}^{\mu\nu\rho\sigma}$ is the Lichnerowicz operator, defined as
\begin{align}
    {\mathcal E}_{\mu\nu}^{~~~\rho\sigma} M_{\rho\sigma} = &-\frac12\left(\Box M_{\mu\nu} - \partial_\mu\partial^\lambda M_{\lambda\nu} - \partial_{\nu}\partial^\lambda M_{\lambda\mu}\right.\nonumber\\
    &\left.+\,\partial_\mu\partial_\nu M -\eta_{\mu\nu}\Box M + \eta_{\mu\nu}\partial_\lambda\partial_\kappa M^{\lambda\kappa}\right),
\end{align}
where $\Box \doteq \eta^{\mu\nu}\partial_\mu\partial_\nu$. Notice the factor of \(1/4\) difference with the conventions of~\cite{Aoki:2016zgp,Manita:2023mnc,LIGOScientific:2025ttj} which results in the field normalisation \(mM_0 = \sqrt{2\rhoDM}\) (instead of twice this value), see \cref{eq:ampnorm}.\footnote{Furthermore, the factor of \(1/2\) difference with respect to the normalisation of~\cite{Babichev:2016hir,Babichev:2016bxi,Marzola:2017lbt} arises from the latter's definition of the stress-energy tensor as \(T_{\mu\nu} \doteq - \delta(\sqrt{|g|}{\cal L}_\mathrm{matter})/\sqrt{|g|}\delta g^{\mu\nu}\), without the factor \(2\) as is more common in the General Relativity literature.}

%%%%%%%%%%%%%%%%%%%%%%%%%%%%%%%%%%%%%%%%%%%%
\section{Derivation of the detector strains}
\label{app:strains}

We provide here more details about the derivation of the strains for the spin-1 and spin-2 cases, as expressed by \cref{eq:strainspin1ts} and \cref{eq:strainspin2ts}.

%%%%%%%%%%
\paragraph{Response to the spin-1 signal.}
Let us begin once again from the acceleration that drives the motion of a detector mirror placed along the unit vector \(\hat{x}\), as in \cref{eq:eomspin1}:
\begin{align}
    \frac{d^2x}{dt^2}=-\epsilon_D e \frac{Q_D}{M} \hat{x}_i \dot{A}^i,
\end{align}
where $M$ is the mirror mass. We can describe the unperturbed positions of the input and end mirrors as
\begin{align}\label{eq:inputandendmirrors}
    x_i(t) = -\frac{L}{2}+\delta x_-(t), \quad x_e(t) = \frac{L}{2}+\delta x_+(t).
\end{align}
The perturbative solution to \cref{eq:eomspin1} provides the displacements in the mirrors
\begin{align}\label{eq:displacementsspin1}
    \delta x_\pm = - \epsilon_D e\frac{Q_D}{M} \sum_{a=1}^N \frac{1}{\omega_a} A_0^a \varepsilon_i^a \hat{x}^i \sin\left(\omega_a t-\vk_a\cdot\vx+\gamma_a\right).
\end{align}
These displacements lead to a phase shift $\phi(t,\vx)$ of the laser in the detector of length $L$ given by
\begin{align}
    \phi(t,\vx)=2\pi\nu(t-T_r)+\phi_0
\end{align}
where $\nu$ is the laser frequency and $T_r$ is the round-trip time in the detector arm
\begin{align}
    T_r = -x_i(t)+2 x_e(t-L)-x_i(t-2L),
\end{align}
which can in turn be written in terms of the displacements as
\begin{align}
    T_r&=\delta L_{\rm t} +\delta L_{\rm s}, \nonumber\\
    \delta L_{\rm t}&\doteq -\delta x_-(t)+2\delta x_-(t-L)-\delta x_-(t-2L), \nonumber\\
    \delta L_{\rm s}&\doteq 2\left[\delta x_+(t-L)-\delta x_-(t-L)\right].
\end{align}
The contribution $\delta L_{\rm t}$ describes the displacement of mirrors during the round-trip of the laser, known as the finite-time travelling effect, while $\delta L_{\rm s}$ represents the difference in spatial displacement between the input and end mirrors. In terms of the displacements given in \cref{eq:displacementsspin1}, we have
\begin{align}
    \delta L_{\rm t} &= -4 \epsilon_D e\frac{Q_D}{M} \sum_{a=1}^N \frac{1}{\omega_a} \sin^2\left(\frac{\omega_a L}{2}\right) A_0^a \varepsilon_i^a \hat{x}^i  \sin\left(\omega_a (t-L)+\gamma_a\right),\nonumber\\
    \delta L_{\rm s} &= 2\epsilon_D e\frac{Q_D L}{M} \sum_{a=1}^N v_a A_0^a \varepsilon_i^a \hat{x}^i \hat{k}_j \hat{x}^j \cos\left(\omega_a (t-L)+\gamma_a\right).
\end{align}

With these expressions at hand we see that the ratio between the time and space contributions is
\begin{align}
    \frac{\delta L_{\rm t}}{\delta L_{\rm s}} \sim \frac{mL}{2v} \simeq 12.56 \left(\frac{m}{2\pi 100 {\rm Hz}}\right)\left(\frac{L}{4{\rm km}}\right)\left(\frac{200{\rm km/s}}{v}\right).
\end{align}
Thence, the $\delta L_{\rm t}$ dominates for $f \gtrsim 9.20\,\mathrm{Hz}$ and for the LVK band, the response function is dominated by
\begin{align}
    \delta L_{\rm t} \propto \hat{x}_i e^i_\lambda.
\end{align}
This is the dominant terms associated to the response function \cref{eq:responsespin1}, whereas for the space contribution the response is instead given by \cref{eq:responsespin2}. Finally, the strain in gravitational wave detectors is computed from the phase shift $\phi(t,{\bf x})$ as
\begin{align}\label{eq:straindef}
    h=\frac{1}{2\pi\nu}\frac{\phi(t,{\bf x})-\phi(t,{\bf y})}{2L},
\end{align}
which, after some easy algebra, gives our \cref{eq:strainspin1ts}.

The next step is to perform the averages over time, velocities, directions and polarisations. In order to average the effect of the distribution of velocities in \cref{eq:MBPDF} for the frequencies $\omega_a$ we can write
\begin{align}
    \left<h\right>_v=\int_0^\infty dv \, {\rm PDF}(v) \, h(\omega_a(v),v),
\end{align}
where $\left<.\right>_v$ denotes average over velocities. By expanding $h(\omega_a)$ around $m$ as
\begin{align}
  h(\omega_a,v_a)\simeq h(\omega,v_a)+\frac{m}{2} \frac{\partial h(\omega,v_a)}{\partial \omega} v_a^2,
\end{align}
where $\omega=m$, and using
\begin{align}
    \left<v_a^2\right>_v=\int_0^\infty dv \, {\rm PDF}(v) \, v^2 = \frac{3}{2}v_0^2,
\end{align}
we find
\begin{align}\label{eq:whywecanuseomega}
    \left<h\right>_v\simeq h(\omega,\sqrt{3/2}v_0)+\frac{3}{4} m v_0^2 \frac{\partial h(\omega,\sqrt{3/2}v_0)}{\partial \omega}.
\end{align}
For the remaining of this section, we keep solely the leading order term, where the strain is monochromatic at $\omega=m$ and $v_a$ is averaged.

Next, the strain averaged over time satisfies $\left<h^2\right>_t=\left<h_{\rm t}^2\right>_t+\left<h_{\rm s}^2\right>_t$~\cite{Morisaki:2020gui}, such that
\begin{align}\label{eq:timeaveragedstrainspin1}
    \left< h^2 \right>_{v,t} = \frac{1}{2}\left[ \left(2\frac{\epsilon_D e}{\omega L}\frac{Q_D}{M}\sin^2\left(\frac{\omega L}{2}\right)A_0 \varepsilon_i D^i\right)^2 + \left(\epsilon_D e \frac{Q_D}{M} \sqrt{\frac{3}{2}}v_0 A_0 2 \varepsilon_i D^{ij} \hat{k}_j\right)^2\right].
\end{align}
The averages over polarisation and direction can be performed without loss of generality within our chosen longitudinal setup $\varepsilon_i=\hat{k}_i$, where the isotropic average over the sphere leads to
\begin{align}
    \left<\hat{k}_i\hat{k}_j\right>_k = \frac{1}{3}\delta_{ij}, \quad \left<\hat{k}_i\hat{k}_j \hat{k}_k \hat{k}_l\right>_k = \frac{1}{15}(\delta_{ij}\delta_{kl}+\delta_{ik}\delta_{jl}+\delta_{il}\delta_{jk}).
\end{align}
Using that $D_{i}D^{i}=2$ and $D_{ij}D^{ij}=1/2$, we find
\begin{align}
    \left<\varepsilon_i D^{ij} \hat{k}_j \varepsilon_l D^{lm} \hat{k}_m\right>_k=\frac{1}{15}, \quad \left<\varepsilon_i D^i \varepsilon_j D^j\right>_k=\frac{2}{3}.
\end{align}
Putting all of this together we recover our final averaged expression for the strain, \cref{eq:avrstrainspin1}, which numerically reads
\begin{align}
    \sqrt{\left<h^2\right>} \simeq 4.45 \times 10^{-23} \left(\frac{\epsilon_D}{10^{-22}}\right)\left(\frac{100{\rm Hz}}{f}\right) \left[8.79\times 10^{-6}\left(\frac{f}{100 {\rm Hz}}\right)^2+1.76 \times 10^{-7}\right]^{\frac{1}{2}}.
\end{align}

%%%%%%%%%%
\paragraph{Response to the spin-2 signal.}
In this case we start from the acceleration equation \cref{eq:geodesiceq2}, which we repeat here:
\begin{align}
    \frac{d^2x}{dt^2}=\frac{\alpha x}{M_{\rm Pl}}\ddot{M}_{jk} \hat{x}^j \hat{x}^k.
\end{align}
Considering the input and end mirrors located at the positions given by \cref{eq:inputandendmirrors}, we find that the acceleration equation is solved by
\begin{align}
    \delta x_\pm =\pm \frac{\alpha L}{2 M_{\rm Pl}} M_{jk} \hat{x}^j \hat{x}^k.
\end{align}
It is worth noting that in the spin-2 case the mirrors move in opposite directions, unlike the case of the spin-1 ultra-light dark matter field. Similarly to the approach carried out for spin 1, we find
\begin{align}
    \delta L_{\rm s} &= \frac{2\alpha L}{M_{\rm Pl}}\hat{x}^i \hat{x}^j \sum_{a=1}^N M_0^a\varepsilon_{ij}^a \cos\left(\omega_a (t-L)+\gamma_a\right), \nonumber\\
    \delta L_{\rm t} &= -\frac{2\alpha L}{M_{\rm Pl}}\hat{x}^i \hat{x}^j \sum_{a=1}^N \sin^2\left(\frac{\omega_a L}{2}\right) M_0^a\varepsilon_{ij}^a \cos\left(\omega_a (t-L)+\gamma_a\right).
\end{align}
The ratio between the time and space contributions is given by
\begin{align}
    \frac{\delta L_{\rm t}}{\delta L_{\rm s}} \simeq \sin^2\left(\frac{mL}{2}\right) \simeq 1.75 \times 10^{-5}\left(\frac{m}{2\pi 100 {\rm Hz}}\right)^2 \left(\frac{L}{4{\rm km}}\right)^2,
\end{align}
which means that in the LVK band the space term will be strongly dominant as expected.

Recasting the results in terms of strain we recover our \cref{eq:strainspin2ts} and, once again following the same steps performed in the spin-1 case, we perform the average over velocities and keep the leading order term in \cref{eq:whywecanuseomega} as well. The intermediate average over time and velocity is given by
\begin{align}\label{eq:timeaveragedstrainspin2}
    \left< h^2 \right>_{v,t} = \frac{2\alpha^2}{M^2_\mathrm{Pl}}\left[ \left(\sin^2\left(\frac{\omega L}{2}\right)D^{ij} M_0 \varepsilon_{ij}\right)^2 + \left(D^{ij} M_0 \varepsilon_{ij}\right)^2\right].
\end{align}
To perform the last step, the averages over polarisation and direction, we first note that the set of matrices given in \cref{eq:Ybasis} is a set of symmetric, traceless, and orthonormal entities. In other words, $\cY_{ij}^\kappa$, where $\kappa\in \{+,\times,L,R,S\}$, form an orthonormal basis of the space of symmetric, traceless $3\times 3$ tensors. The most general isotropic projector onto such space is constructed from Kronecker deltas as
\begin{align}
    P_{ijkl} = \sum_\kappa \cY_{ij}^\kappa \cY_{kl}^\kappa = \frac{1}{2}(\delta_{ik}\delta_{jl}+\delta_{il}\delta_{jk})-\frac{1}{3}\delta_{ij}\delta_{kl}.
\end{align}
As a result, the averaged polarisation tensor is given by
\begin{align}
    \left<\varepsilon_{ij} \varepsilon_{kl}\right>_\varepsilon = \frac{1}{5}\sum_\kappa \cY_{ij}^\kappa \cY_{kl}^\kappa = \frac{1}{5}\left[\frac{1}{2}(\delta_{ik}\delta_{jl}+\delta_{il}\delta_{jk})-\frac{1}{3}\delta_{ij} \delta_{kl}\right],
\end{align}
such that
\begin{align}
    \left<D^{ij} \varepsilon_{ij} D^{kl}\varepsilon_{kl}\right>_\varepsilon=\frac{1}{10}.
\end{align}
With this last expression at hand we directly recover our final, averaged spin-2 ultra-light dark matter strain quoted in \cref{eq:avrstrainspin2}.

%%%%%%%%%%%%%%%%%%%%%%%%%%%%%%%%%%%%%%%%%%%%
\section{Peaks amplitudes}
\label{app:peaks}

\allowdisplaybreaks

In this appendix we collect the coefficients that determine the relative amplitudes of the spectral peaks in \cref{eq:dep_s1t}, \cref{eq:dep_s1s} and \cref{eq:dep_s2}. The spin-1 coefficients for the \(\Delta\varepsilon_\mathrm{t}\) term are:

\begin{align*}
    \Delta\varepsilon^\mathrm{t}_{-1}  &= -\left(\varepsilon_p \cos\theta + \varepsilon_r \sin\theta\right)
    \left[ \cos\phi\left(\sin\xi+\cos\xi\right) + \sin\phi\left(\sin\xi-\cos\xi\right)\sin\lambda \right] \\
    &~~~+ \varepsilon_q \left[ \sin\phi\left(\sin\xi+\cos\xi\right) - \cos\phi\left(\sin\xi-\cos\xi\right)\sin\lambda \right], \\
    \nonumber\Delta\varepsilon^\mathrm{t}_{0} &= (\varepsilon_r \cos\theta - \varepsilon_p \sin\theta)\, (\sin\xi-\cos\xi ) \cos\lambda, \\
    \Delta\varepsilon^\mathrm{t}_{1}  &= \left(\varepsilon_p \cos\theta + \varepsilon_r \sin\theta\right)
    \left[ \sin\phi\left(\sin\xi+\cos\xi\right) - \cos\phi\left(\sin\xi-\cos\xi\right)\sin\lambda \right] \\
    &~~~+ \varepsilon_q \left[ \cos\phi\left(\sin\xi+\cos\xi\right) + \sin\phi\left(\sin\xi-\cos\xi\right)\sin\lambda \right].
\end{align*}

The coefficients of the \(\Delta\varepsilon_\mathrm{s}\) term are: 
\begin{align}
\nonumber \Delta\varepsilon^\mathrm{s}_{-2} =&
-\frac{1}{8}\cos 2\xi\,(3 - \cos 2\lambda)
\Bigl\{
\frac{1}{2}\left[\varepsilon_p \sin 2\theta + \varepsilon_r(1 - \cos 2\theta)\right]\sin 2\phi
+ \varepsilon_q \sin\theta \cos 2\phi
\Bigr\}\\
\nonumber&+ \frac{1}{2}\sin 2\xi\,\sin\lambda
\Bigl\{
\frac{1}{2}\left[\varepsilon_p \sin 2\theta + \varepsilon_r(1 - \cos 2\theta)\right]\cos 2\phi
- \varepsilon_q \sin\theta \sin 2\phi
\Bigr\},\\
\nonumber\Delta\varepsilon^\mathrm{s}_{-1} =&
\frac{1}{4}\cos 2\xi\,\sin 2\lambda
\left[
(\varepsilon_p \cos 2\theta + \varepsilon_r \sin 2\theta)\sin\phi
+ \varepsilon_q \cos\theta \cos\phi
\right] \\
\nonumber&+ \frac{1}{2}\sin 2\xi \cos\lambda
\left[
-(\varepsilon_p \cos 2\theta + \varepsilon_r \sin 2\theta)\cos\phi
+ \varepsilon_q \cos\theta \sin\phi
\right],\\
\nonumber\Delta\varepsilon^\mathrm{s}_{0} =&
- \frac{1}{8} \cos2\xi \cos^2\lambda
\left[
\varepsilon_r\left(1 + 3  \cos2\theta\right) - 3 \varepsilon_p \sin2\theta
\right],\\
\nonumber\Delta\varepsilon^\mathrm{s}_{1} =&
\frac{1}{4}\cos 2\xi \sin 2\lambda
\left[
(\varepsilon_p \cos 2\theta + \varepsilon_r \sin 2\theta)\cos\phi
- \varepsilon_q \sin\phi \cos\theta
\right] \\
\nonumber&+ \frac{1}{2}\sin 2\xi \cos\lambda
\left[
(\varepsilon_p \cos 2\theta + \varepsilon_r \sin 2\theta)\sin\phi
+ \varepsilon_q \cos\theta \cos\phi
\right],\\
\nonumber\Delta\varepsilon^\mathrm{s}_{2} =&
-\frac{1}{8}\cos 2\xi(3 - \cos 2\lambda)
\Bigl\{
\frac{1}{2}[\varepsilon_p \sin 2\theta + \varepsilon_r(1 - \cos 2\theta)]\cos 2\phi
- \varepsilon_q \sin\theta \sin 2\phi
\Bigr\} \\
\nonumber&+ \frac{1}{2}\sin 2\xi \sin\lambda
\Bigl\{
-\frac{1}{2}[\varepsilon_p \sin 2\theta + \varepsilon_r(1 - \cos 2\theta)]\sin 2\phi
- \varepsilon_q \sin\theta \cos 2\phi
\Bigr\}.
\end{align}

Finally, the spin-2 peak amplitude coefficients of \(\Delta\varepsilon\) are given by:
\begin{align}
\nonumber\Delta\varepsilon_{-2}
=& 
\frac{(-3 + \cos 2\lambda)\cos 2\xi}{16\sqrt{2}}
\Bigl\{
4 (\varepsilon_x \cos\theta + \varepsilon_L \sin\theta)\cos 2\phi
+ \Bigl[\varepsilon_{+}(3+\cos 2\theta)\\
\nonumber & + 2\varepsilon_R \sin 2\theta + \sqrt{3}\,\varepsilon_S (1 - \cos 2\theta)\Bigl]\sin 2\phi
\Bigr\} + \frac{\sin\lambda\sin 2\xi}{4\sqrt{2}}
\Bigl\{- 4(\varepsilon_x \cos\theta + \varepsilon_L \sin\theta)\sin 2\phi\\
\nonumber &+
\left[\varepsilon_{+}(3+\cos 2\theta) + 2\varepsilon_R \sin 2\theta + \sqrt{3}\,\varepsilon_S (1 - \cos 2\theta)\right]\cos 2\phi
\Bigr\},
\\
\nonumber\Delta\varepsilon_{-1}=&
\frac{\cos 2\xi\,\sin 2\lambda}{2\sqrt{2}}
\Bigl[
(-\varepsilon_x \sin\theta + \varepsilon_L \cos\theta)\cos\phi
+ \Bigl(\varepsilon_R \cos 2\theta - \frac{\varepsilon_{+} - \sqrt{3}\,\varepsilon_S}{2}\sin 2\theta\Bigl)\sin\phi
\Bigl] \\
\nonumber&+ \frac{\cos\lambda\,\sin 2\xi}{\sqrt{2}}
\Bigl[
(-\varepsilon_x \sin\theta + \varepsilon_L \cos\theta)\sin\phi
- \Bigl(\varepsilon_R \cos 2\theta - \frac{\varepsilon_{+} - \sqrt{3}\,\varepsilon_S}{2}\sin 2\theta\Bigl)\cos\phi
\Bigl],
\\
\nonumber\Delta\varepsilon_{0}
=& \frac{\cos^{2}\lambda\cos 2\xi}{8\sqrt{2}}
\left[
-3\varepsilon_{+}
-\sqrt{3}\,\varepsilon_{S}
+3(\varepsilon_{+}-\sqrt{3}\,\varepsilon_{S})\cos 2\theta
+6\varepsilon_{R}\sin 2\theta
\right],
\\
\nonumber\Delta\varepsilon_{+1}
=&
\frac{\cos 2\xi\,\sin 2\lambda}{2\sqrt{2}}
\Bigl[ -(-\varepsilon_x \sin\theta + \varepsilon_L \cos\theta)\sin\phi+
\Bigl(\varepsilon_R \cos 2\theta - \frac{\varepsilon_{+} - \sqrt{3}\,\varepsilon_S}{2}\sin 2\theta\Bigl)\cos\phi
\Bigl] \\
\nonumber&+ \frac{\cos\lambda\,\sin 2\xi}{\sqrt{2}}
\Bigl[
(-\varepsilon_x \sin\theta + \varepsilon_L \cos\theta)\cos\phi
+ \Bigl(\varepsilon_R \cos 2\theta - \frac{\varepsilon_{+} - \sqrt{3}\,\varepsilon_S}{2}\sin 2\theta\Bigl)\sin\phi
\Bigl],
\\
\nonumber\Delta\varepsilon_{+2}
=& 
\frac{(-3 + \cos 2\lambda)\cos 2\xi}{16\sqrt{2}}
\Bigl\{- 4(\varepsilon_x \cos\theta + \varepsilon_L \sin\theta)\sin 2\phi+
\Bigl[\varepsilon_{+}(3 + \cos 2\theta)\\
\nonumber& + 2\varepsilon_R \sin 2\theta + \sqrt{3}\,\varepsilon_S (1 - \cos 2\theta)\Bigl]\cos 2\phi
\Bigr\}+ \frac{\sin\lambda\sin 2\xi}{4\sqrt{2}}
\Bigl\{
-4 (\varepsilon_x \cos\theta + \varepsilon_L \sin\theta)\cos 2\phi\\
\nonumber &- \Bigl[\varepsilon_{+}(3 + \cos 2\theta) + 2\varepsilon_R \sin 2\theta + \sqrt{3}\,\varepsilon_S (1 - \cos 2\theta)\Bigl]\sin 2\phi
\Bigr\}.
\end{align}

%%%%%%%%%%%%%%%%%%%%%%%%%%%%%%%%%%%%%%%%%%%%
\section{Details on the \bsdexcess method}
\label{app:bsd}

In order to calculate the theoretical sensitivity of the \bsdexcess method, we follow the approach carried out in~\cite{PhysRevD.90.042002} with the correction made by~\cite{Palomba2025_peakmap_sensitivity}.  
We aim at identifying time-frequency bins whose normalised power exceeds a predefined threshold. The  threshold crossings are then mapped into a frequency parameter space, where a signal appears as a statistically significant accumulation of counts. 

We make use of the probability density of the spectrum in the presence of a signal
\begin{align}
    p(x;\lambda)={\rm e}^{-x-\frac{\lambda}{2}}I_0(\sqrt{2x\lambda}),
\end{align}
where $I_0$ is the modified Bessel function of zeroth order, $\lambda$ is the signal spectral amplitude, and $x$ is the equalised power in a single time-frequency bin of the peakmap. Then the probability of selecting a local maxima above a chosen
threshold $\theta$ is given by
\begin{align}
    p_\lambda=\int_\theta^\infty p(x;\lambda)\left(\int_0^x {\rm e}^{-{x}'}d{x}'\right)^2 dx.
\end{align}
In the limit of small signals $\lambda\ll1$, the probability of selecting a noise peak above the threshold $\theta$ is given by $p_0$, 
while the probability of a signal being present is given by $ p_\lambda\simeq p_0+\lambda p_1$, where $p_0$ and $p_1$ are provided in~\cite{Palomba2025_peakmap_sensitivity}.
We finally require that the probability of selecting a candidate as a function of the signal spectral amplitude $\lambda$ is equal to the confidence level $\Gamma$.  
Under the Gaussian approximation to the probability distribution of the map, 
we find the approximate  
solution for $\lambda$~\cite{PhysRevD.90.042002}
\begin{align}
    \lambda_{\rm min}(f_j) \simeq \sqrt{\frac{p_0(1-p_0)}{N_{\rm seg}(f_j) p_1^2}}\left[{\rm CR}_\mathrm{thr}-\sqrt{2}{\rm erfc}^{-1}(2\Gamma)\right],
\end{align}
where $N_{\rm seg}(f_j)$ is the number of FFTs in the frequency bin $j$, and ${\rm CR}_\mathrm{thr}$ is the critical ratio threshold used for the first level candidates selection.
The sensitivity $\lambda_{\rm min}(f_j)$ can be translated to the ultra-light dark matter strain via \begin{align}
    |\tilde{h}_{\rm min}(f_j)|&=\left[\frac{\lambda_{\rm min}(f_j)}{4}T_{\rm seg}(f_j){S_n}(f_j)\right]^{\frac{1}{2}},
\end{align}
where ${S_n(f_j)}$ is the detector one-sided PSD, 
and $\tilde{h}(f_j)$ is related to $\sqrt{\left<h^2\right>}$ via \cref{eq:hfourier}. 
As a result we find 
\begin{align}
    \sqrt{\left<h^2\right>}_{\rm min} = \left[\frac{\pi}{2.43}\frac{1}{2}\frac{ S_n(f_j)}{ T_{\rm seg}(f_j)}\right]^{\frac{1}{2}}\left(\frac{p_0(1-p_0)}{N_{\rm seg}(f_j) p_1^2}\right)^{\frac{1}{4}} \sqrt{{\rm CR}_{\rm thr}-\sqrt{2} {\rm erfc}^{-1}(2\Gamma)},
\end{align}
which is reported in a simplified form in \cref{eq:bsdsensitivity} in the main text and we can recognise $\mathcal{B}=3.44$ for ${\rm CR}_{\rm thr}=5$ and a confidence level $\Gamma=0.95$.

\bibliographystyle{JHEP}
\bibliography{main}

\end{document}